\documentclass[lettersize,journal]{IEEEtran}
\usepackage{amsmath,amsfonts}
\usepackage{algorithmic}
\usepackage{algorithm}
\usepackage{array}
\usepackage[caption=false,font=normalsize,labelfont=sf,textfont=sf]{subfig}
\usepackage{textcomp}
\usepackage{stfloats}
\usepackage{url}
\usepackage{verbatim}
\usepackage{graphicx}
\usepackage{cite}
\usepackage{xcolor}  

\usepackage{amsmath}    

\usepackage{diagbox}
\usepackage{booktabs} 
\usepackage{makecell}
\usepackage{multirow} 

\usepackage{amssymb}
\makeatletter

\newcommand{\Rmnum}[1]{\expandafter\@slowromancap\romannumeral #1@}
\makeatother

\usepackage{etoolbox}
\makeatletter
\patchcmd{\@makecaption}
  {\scshape}
  {}
  {}
  {}
\makeatletter
\patchcmd{\@makecaption}   
  {\\}
  {.\ }
  {}
  {}
\makeatletter

\usepackage[
bookmarks=true,
colorlinks=true,
linkcolor=blue,
urlcolor=blue,
citecolor=blue,
pdftex,
linktocpage=true,   
hyperindex=true
]{hyperref}

\begin{document}

\title{DINVMark: A Deep Invertible Network for Video Watermarking}

\author{Jianbin Ji, Dawen Xu,~\IEEEmembership{{Senior Member,~IEEE}}, Li Dong,~\IEEEmembership{Member,~IEEE}, Lin Yang, Songhan He
\thanks{This work is supported by the National Natural Science Foundation of
China (62471269, 62071267, 62171244) and Ningbo Natural Science Foundation (2023J022, 2024J210). 
(Corresponding authors: Dawen Xu; Li Dong)

J. Ji, L. Dong, L. Yang and S. He are with the Faculty of Electrical Engineering and Computer Science, Ningbo University, Ningbo, 315211, China.
D. Xu is with the School of Cyber Science and Engineering, Ningbo University
of Technology, 315211, Ningbo, China (e-mail: dawenxu@126.com).}}

\markboth{IEEE TRANSACTIONS ON MULTIMEDIA}%
{Shell \MakeLowercase{\textit{et al.}}: A Sample Article Using IEEEtran.cls for IEEE Journals}


\maketitle

\begin{abstract}
	\textbf{{With the wide spread of video, video watermarking has become increasingly crucial for copyright protection and content authentication. However, video watermarking still faces numerous challenges. For example, existing methods typically have shortcomings in terms of watermarking capacity and robustness, and there is a lack of specialized {noise layer for High Efficiency Video Coding(HEVC) compression.} To address these issues, this paper introduces a Deep Invertible Network for Video watermarking (DINVMark) and designs a {noise layer} to simulate HEVC compression. This approach not only increases watermarking capacity but also enhances robustness. DINVMark employs an Invertible Neural Network (INN), where the encoder and decoder share the same network structure for both watermark embedding and extraction. This shared architecture ensures close coupling between the encoder and decoder, thereby improving the accuracy of the watermark extraction process. Experimental results demonstrate that the proposed scheme significantly enhances watermark robustness, preserves video quality, and substantially increases watermark embedding capacity. }}
\end{abstract}


\section{Introduction}
\IEEEPARstart{I}{n} the information age, protecting data and ensuring secure communication have become critical priorities. Information hiding technologies, such as steganography and digital watermarking \cite{ch1,ch2}, play a key role in this context. Steganography focuses on embedding secret information within digital media in a way that prevents detection during transmission, making it ideal for covert communication \cite{01,02,03}. In contrast, digital watermarking prioritizes robustness, ensuring that the embedded watermark can be reliably extracted even after operations like compression or channel processing. This robustness enables watermarking to support applications such as copyright protection, content authenticity verification, and the prevention of unauthorized copying. It also provides valuable evidence in copyright disputes. With recent advances in deep learning, the performance of watermark embedding and extraction has significantly improved, making digital watermarking more resilient to attacks while preserving the visual quality of multimedia content. As a result, digital watermarking has become a vital tool for modern information security.

Digital watermarking is categorized into visible and invisible watermark, based on whether the watermark is visible to the naked eye. Visible watermark is the direct embedding of perceptible watermark into video, which is more likely to be retained under non-human attack, however, the visual quality of the carrier will inevitably be affected by visible watermark, and the watermark will be lost in the case of such attacks as cropping. Invisible watermark entails embedding watermark without degrading video visual quality, and it typically does not affect the original content. Among them, invisible watermark can be further classified into spatial domain watermarking, transformed domain watermarking and compressed domain watermarking according to the different embedding domains. Spatial domain watermarking schemes \cite{04} embeds the watermark directly into the video pixel domain, offering high computational efficiency and fast processing speeds. However, it is usually weak in resisting geometric attacks. Therefore, researchers have turned their attention to transformed domain watermarking  \cite{05, 06, 07}. Transformed domain watermarking schemes involves embedding the watermark in the transform domain coefficients, which are obtained by Discrete Cosine Transform (DCT), Discrete Fourier Transform (DFT), Discrete Wavelet Transform (DWT), and Dual Tree-Complex Wavelet Transform (DT CWT). These methods are better at resisting common attacks like recompression and geometric distortions, though they tend to be more complex and effective only against specific types of attacks. Compressed domain watermarking schemes \cite{08, 09, 10, 11}, which embeds the watermark in the encoded bitstream of standard encoders like H.264/AVC or HEVC, is compatible with the video coding process and performs well in terms of embedding efficiency, but it is less robust to recompression attacks.

\begin{figure}[!t]
\centering
\scalebox{1.35}{
\includegraphics[width=2.5in]{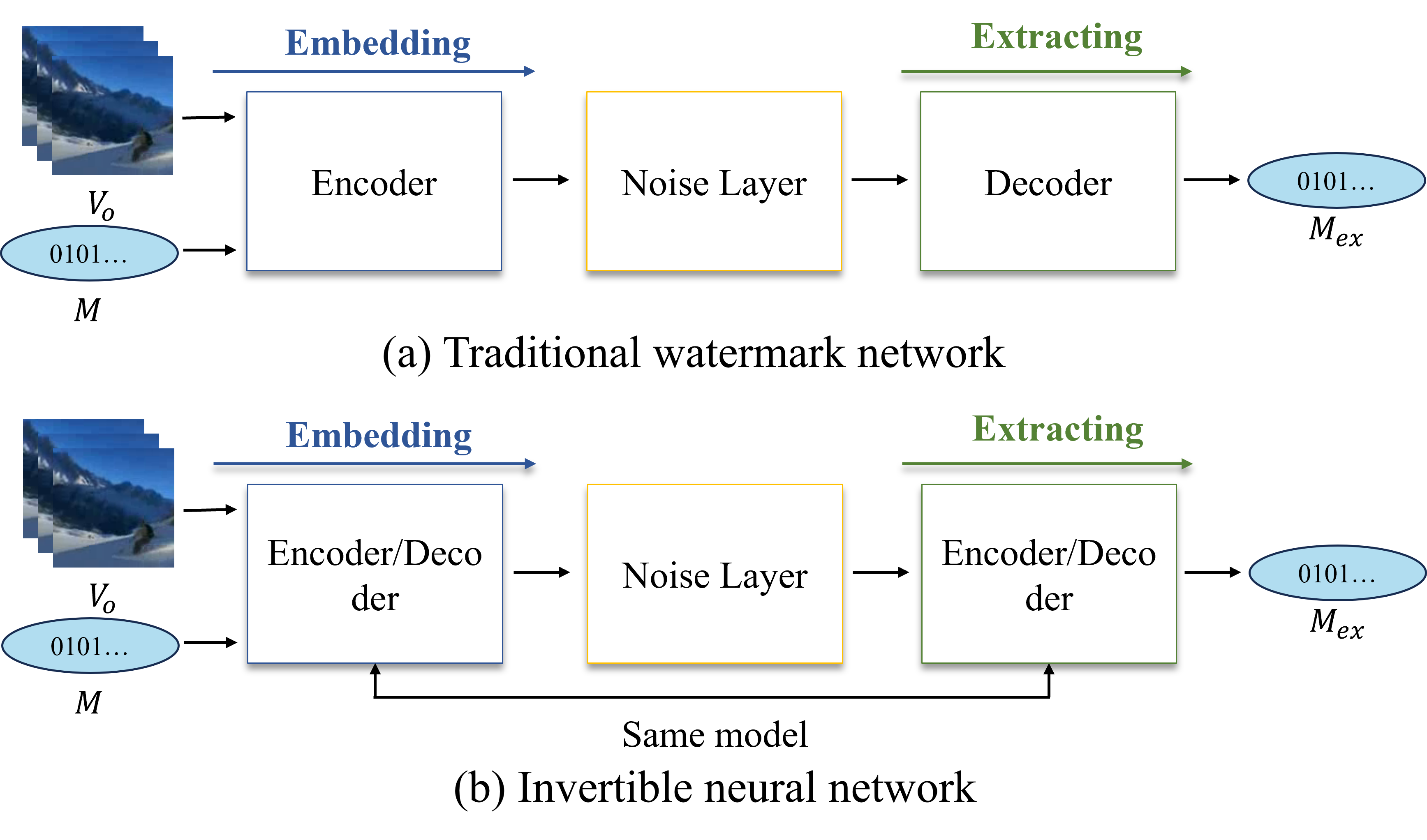}
}
\caption{The basic difference between traditional watermark network and invertible neural network}
\label{innAndTradition}
\end{figure}

Recent breakthroughs in deep learning, particularly the use of Convolutional Neural Networks (CNNs) and Generative Adversarial Networks (GANs), have significantly enhanced the performance of digital watermarking, allowing watermarking algorithms to learn complex patterns in video content, enabling the watermark to be embedded in optimal locations for both transparency and robustness. Typically, deep learning-based video watermarking frameworks consist of an encoder, which embeds the watermark into the video, and a decoder, which effectively extracts the watermark. To enhance the robustness of watermarked video transmission over lossy channels, a differentiable distortion layer is typically incorporated between the encoder and decoder during training, to better simulate the actual lossy channel transmission, as shown in Fig. \ref{innAndTradition}(a). While much progress has been made in image watermarking with networks such as HiDDeN \cite{hidden}, StegaStamp \cite{StegaStamp}, and ReDMark \cite{redmark}, video watermarking remains a developing field with many challenges. One key issue is that many existing networks, originally designed for image watermarking, do not fully leverage the temporal correlation in video data, leading to suboptimal performance for video. Additionally, although the encoder and decoder of such networks are trained simultaneously, the lack of strong coupling between them can result in the encoder embedding features that the decoder struggles to extract, thereby limiting the performance of the model. Another limitation is that current video watermarking can only be embedded with small capacity, meaning cannot encompass various data such as author information, copyright declarations, and version numbers simultaneously. {The advantages of the proposed method, in comparison with other existing watermarking methods, are summarized in Table \ref{advantages}. A checkmark “\checkmark” signiﬁes the incorporation of a speciﬁc performance-enhancing aspect in the method. In response to the aforementioned issues, this paper proposes the DINVMark video watermarking network, as shown in Fig. \ref{innAndTradition}(b). This network enhances the robustness of the watermark, maintains video quality, and increases watermarking capacity. DINVMark demonstrates strong robustness against various attacks, including compression and noise, making it highly suitable for protecting the intellectual property rights. By embedding unique watermark into video frames, the proposed method can effectively provide evidence. Additionally, the lightweight nature of the proposed method allows for efficient implementation, making it practical for applications. }

\begin{table*}
\centering
\caption{{Advantages of the proposed method compared with other watermarking methods}}
\label{advantages}
\begin{tabular}{lcccc}
\toprule
{Framework} & {Embedding method} & {Visual Quality} & {Robustness for video compression} & {Embedding Capacity} \\
\midrule
{HiDDeN[14]} & {Deep learning watermarking} & {\checkmark} &  & \\
{DT-CWT[8]} & {Transformed domain watermarking} & {\checkmark} & {\checkmark} & \\
{DVMark[19]} & {Deep learning watermarking} & {\checkmark} & {\checkmark} & \\
{REVMark[20]} & {Deep learning watermarking} & {\checkmark} & {\checkmark} & \\
{DINVMark(ours)} & {Deep learning watermarking} & {\checkmark} & {\checkmark} & {\checkmark} \\
\bottomrule
\end{tabular}
\end{table*}

\subsubsection*{The main contributions of this paper are as follows}
\begin{itemize}
\item{A differentiable video compression simulator is designed to achieve robustness against video compression in video watermarking.}
\item{INN is employed for watermark embedding and extraction, avoiding the issue of poor coupling between the encoder and decoder, and thus increasing the capacity of the watermark.}
\item{Compared to existing schemes, DINVMark achieves simpler network structure and lower complexity, and DINVMark demonstrates better robustness against various attacks such as video compression, frame swap, and frame drop.}
\end{itemize}

The rest of this paper is organized as follows. The related work is introduced in Section \Rmnum{2}. In Section \Rmnum{3}, the proposed scheme is presented. Experimental results are discussed and analyzed in Section \Rmnum{4}, followed by the conclusions and future work in the last section.

\section{RELATED WORK}
\subsection{Video Watermarking}
Digital watermarking, characterized by robustness and diversity, serves the functions of anti-counterfeiting verification and traceability in practical fields such as copyright protection and content authentication. Chen \textit{et al.} \cite{211} proposed a fast video watermarking method that embeds watermark into selected frame pairs of videos by utilizing Zernike moments and Singular Value Decomposition (SVD), thereby achieving robust, imperceptible video watermarking. Huan \textit{et al.} \cite{06} discovered the relationships between the sub-bands of the DT CWT and embedded watermark within the established pairs of joint sub-bands. They extracted the watermark from the joint sub-bands based on a coefficient selection strategy, making it effective in resisting various types of attacks. Due to the temporal dimension of videos, embedding watermark often leads to inter-frame distortion and is not robust enough against video compression. Consequently, Zhang \textit{et al.} \cite{221} proposed RIVAGAN, which employs a custom attention-based mechanism to embed watermark and utilizes two separate adversarial networks to enhance video quality and robustness. Luo \textit{et al.} \cite{222} incorporates a novel multiscale design in both the encoder and decoder, enabling the embedding of watermark across both temporal and spatial dimensions, thereby further enhancing robustness. Zhang \textit{et al.} \cite{223} proposed a temporal-associated feature extraction block (TAsBlock) and a differentiable video compression simulator, enabling resistance against H.264/AVC compression.

\subsection{Invertible Neural Network and Data Hiding}
INN is a generative network based on normalizing flow, which has attracted significant attention since its proposal by Dinh \textit{et al.} \cite{231,232}. To date, INN has been applied to a variety of visual domains, such as image rescaling \cite{2331}, image restoration \cite{2341,2342}, and image-to-image translation \cite{2351}. INN learns a stable invertible mapping between one space and a latent space, this feature is similar to the process of embedding and extracting information. Therefore, it is also widely used in the field of steganography and watermarking. Jing \textit{et al.} \cite{2361} were the first to apply INN to the task of hiding image within an image, transforming the image into the wavelet domain for concealment, while Lu \textit{et al.} \cite{2362} successfully demonstrated the capability to hide multiple different images within a single image. It is essential to consider the impact of external noise attacks in real-world scenarios. To address issue, Xu \textit{et al.} \cite{2363} introduced a distortion layer based on previous work and achieved improved results. Shen \textit{et al.} \cite{2364} applied INN to the task of hiding one video within another video and utilized a differentiable distortion layer to simulate video compression. Mou \textit{et al.} \cite{2365} further increased the number of videos that could be hidden, enabling the concealment of up to seven videos within a single video, and allowed for the extraction of a corresponding secret video from the stego video using a specific key. In the field of watermarking, Xu \textit{et al.} \cite{2371} first applied INN to robust blind image watermarking and designed a bit message normalization module for compressing messages to embed the watermark. Ma \textit{et al.} \cite{2372} combined invertible and non-invertible mechanisms and designed a non-invertible attention-based module, a diffusion and extraction module, and a fusion and split module to enhance the robustness of the watermark. Fang \textit{et al.} \cite{2373} utilized INN as a distortion layer to ensure robustness against black-box distortions. However, existing video watermarking schemes have limitations in terms of watermark capacity, making it impossible to simultaneously embed multiple information, such as author information, copyright declarations, and version numbers. Moreover, these watermarking schemes exhibit weak robustness when faced with HEVC compression attacks. Therefore, in this paper, INN is employed as a distortion layer to simulate HEVC compression and as the primary framework for high-capacity watermark embedding and extraction.

\section{PROPOSED FRAMEWORK}
\begin{figure*}[!t]
\centering
\includegraphics[width=\textwidth]{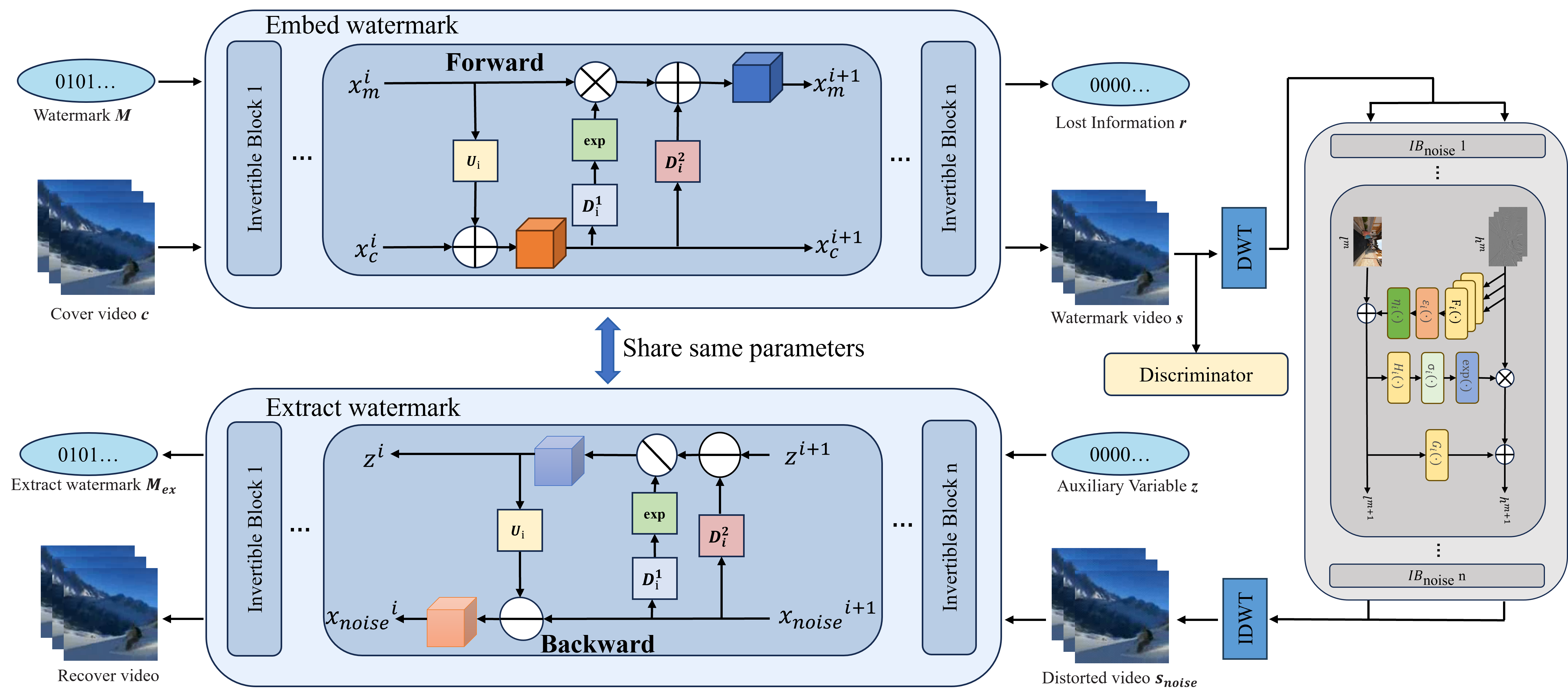}
\caption{{Architecture of the DINVMark network. The invertible neural network serves as both encoder and decoder. During watermark embedding, the encoder integrates the input watermark into the cover video through multiple invertible neural blocks. The noise layer simulates video compression, generates attack videos, and guides the decoder to accurately extract the watermark.}}
\label{framework}
\end{figure*}

An end-to-end trainable robust watermarking model is proposed in this paper, designed to embed a larger capacity watermark while ensuring effective extraction even after HEVC compression. The framework of the proposed model is shown in Fig. \ref{framework}. The core of the framework is an encoder/decoder structure based on the INN, which is composed of multiple invertible neural blocks (INB). The invertibility feature of INN allows for efficient forward encoding and backward decoding processes, which are similar to the embedding and extraction processes of video watermarking. Additionally, the network shares the same parameters for forward encoding and backward decoding, thereby increasing the coupling of them. Compared to traditional separate encoder-decoder frameworks, this network demonstrates higher efficiency and performance. 

The embedding and extraction processes are outlined as follows. In the forward encoding process, the cover video $\boldsymbol{\mathit{c}}$ and watermark $\boldsymbol{\mathit{M}}$ are taken as input, and the watermarked video $\boldsymbol{\mathit{s}}$ and the loss information $\boldsymbol{\mathit{r}}$ are output. The watermarked video is then distorted into the distorted video $\boldsymbol{\mathit{s_{\text{noise}}}}$ through a pre-trained differentiable distortion layer. In the backward watermark extraction process, the distorted video and the all-zero vector $\boldsymbol{\mathit{z}}$ are provided as inputs to the decoder, which reconstructs the extracted watermark $\boldsymbol{\mathit{M_{\text{ex}}}}$. Additionally, the framework includes a discriminator, which serves to enhance the video quality, in addition to the encoder/decoder and the distortion layer.

\subsection{Invertible Encoder/Decoder}
Flow-based models consist of two primary components: the forward encoding function 
represented as ${\mathit{f_\theta}}$, and its corresponding inverse function ${\mathit{f_\theta}^{-1}}$, both sharing the same parameter ${\theta}$. During the forward encoding process, the network receives the watermark $\boldsymbol{\mathit{M}}$ and the cover video $\boldsymbol{\mathit{c}} \in \mathbb{R}^{l \times h \times w \times 3}$, and then outputs the loss information $\boldsymbol{\mathit{r}}$ and the watermarked video $\boldsymbol{\mathit{s}} \in \mathbb{R}^{l \times h \times w \times 3}$. In the backward decoding process, the network takes $\boldsymbol{\mathit{s_{\text{noise}}}} \in \mathbb{R}^{l \times h \times w \times 3}$ along with the all-zero vector $\boldsymbol{\mathit{z}}$ as input and decodes them to produce the extracted watermark $\boldsymbol{\mathit{M_{\text{ex}}}}$.

The INN is combined with $\mathit{n}$ INBs, each with the same network structure, as shown in Fig. \ref{framework} for the $\mathit{i}$-th INB. For the forward encoding process, the inputs are $\boldsymbol{\mathit{x_{\text{c}}}^{i}}$ and $\boldsymbol{\mathit{x_{\text{m}}}^{i}}$, and the outputs are $\boldsymbol{\mathit{x_{\text{c}}}^{i+1}}$ and $\boldsymbol{\mathit{x_{\text{m}}}^{i+1}}$, which are denoted as
\begin{equation}
\boldsymbol{x_{c}^{i+1}}=\boldsymbol{x_{c}^{i}}+U_{i}\left(\boldsymbol{x_{m}^{i}}\right),
\end{equation}
\begin{equation}
\boldsymbol{x_m^{i+1}} = \boldsymbol{x_m^i} \otimes \exp\left(D^1_i(\boldsymbol{x_c^{i+1}})\right) + D^2_i(\boldsymbol{x_c^{i+1}}),
\end{equation}
where $\otimes$ represents the dot product operation, the functions $U(\cdot)$ and $D(\cdot)$ represent upsampling and downsampling processes, respectively. They are responsible for sampling the watermark and the video to the same dimension, allowing for integration. After passing through the last INB, we obtain $\boldsymbol{\mathit{x_{\text{c}}}^{n+1}}$ and $\boldsymbol{\mathit{x_{\text{m}}}^{n+1}}$, which represent the watermarked video $\boldsymbol{\mathit{s}}$ and loss information $\boldsymbol{\mathit{r}}$, respectively. 

For the backward decoding process, the information is fed backward starting from the
$\mathit{n}$-th INB. The inputs are the all-zero vector $\boldsymbol{\mathit{z}^{n+1}}$ 
and the distorted video $\boldsymbol{\mathit{x_{\text{noise}}}^{n+1}}$ after passing through the distortion layer. The outputs are $\boldsymbol{\mathit{z}^{n}}$ and $\boldsymbol{\mathit{x_{\text{noise}}^{n}}}$, which are defined as 
\begin{equation}
\boldsymbol{z^n} = \left(\boldsymbol{z^{n+1}} - D^2_n\left(\boldsymbol{x_{\text{noise}}^{n+1}}\right)\right) \otimes \exp\left(-D^1_n\left(\boldsymbol{x_{\text{noise}}^{n+1}}\right)\right),
\end{equation}
\begin{equation}
\boldsymbol{x_{\text{noise}}^n} = \boldsymbol{x_{\text{noise}}^{n+1}} - U_n(\boldsymbol{z^n}).
\end{equation}

After passing through the last INB, the extracted watermark $\boldsymbol{\mathit{M_{\text{ex}}}}$, $\mathit{i.e.}$, $\boldsymbol{\mathit{z}^{1}}$, will be obtained. To achieve blind extraction, which means successfully extracting the watermark without any other prior information, DINVMark utilizes an all-zero vector as auxiliary information during the backward extraction process.

\subsection{Noise Layer}
To enhance the robustness of the watermark, especially against non-differentiable operations such as HEVC compression, this paper designs a differentiable distortion layer to simulate the effects of HEVC compression. The distortion layer also uses an INN as the back-bone network, but with necessary internal structural adjustments to better simulate the impact of HEVC compression on the watermark. To distinguish the distortion layer from the invertible neural blocks in the watermark network, each invertible neural block in the distortion layer is referred to as $\text{DWT\_INB}$. It is important to note that the distortion layer needs to be pre-trained before watermark embedding and extraction network training. With the above design, the watermark is not only resistant to HEVC compression but also maintains robustness against other attacks that may cause changes in the video content. 

DINVMark primarily simulates the effects of HEVC compression through the following two steps. First, the watermarked video is transformed using the DWT to decompose it into independent low-frequency and high-frequency frequency domain signals, achieving a transition from the pixel domain to the transform domain. This transformation converts 
$\boldsymbol{\mathit{s}} \in \mathbb{R}^{l \times h \times w \times 3}$ into two parts: 
$\boldsymbol{\mathit{l}} \in \mathbb{R}^{l \times h \times w \times 3}$ and $\boldsymbol{\mathit{h}} \in \mathbb{R}^{3 \times l \times h \times w \times 3}$, where 
$\boldsymbol{\mathit{l}}$ and $\boldsymbol{\mathit{h}}$ represent the low-frequency and high-frequency signals, respectively. The aforementioned process can be denoted as 
\begin{equation}
\boldsymbol{\mathit{l}}, \boldsymbol{\mathit{h}} = \text{DWT}(\boldsymbol{\mathit{s}})
\end{equation}

Subsequently, the decomposed low-frequency signal $\boldsymbol{\mathit{l}^{i}}$ and 
high-frequency signal $\boldsymbol{\mathit{h}^{i}}$ are simultaneously input into the $\mathit{i}$-th $\text{DWT\_INB}$, generating new $\boldsymbol{\mathit{l}^{i+1}}$ and $\boldsymbol{\mathit{h}^{i+1}}$, which can be expressed as 
\begin{equation}
\boldsymbol{\mathit{l^{i+1}}} = \boldsymbol{\mathit{l^i}} + \eta_{i} \left( \varepsilon_{i} \left( F_{i} \left( \boldsymbol{\mathit{h^i(j)}} \right) \right) \right),
\end{equation}
\begin{equation}
\begin{aligned}
&\boldsymbol{\mathit{h^{i+1}(j)}} = \\ &\boldsymbol{\mathit{h^i(j)}} \otimes \exp\left(\sigma_{i}\left(H_{i}(\boldsymbol{\mathit{l^{i+1}}})\right) \times 2 - 1\right) + G_{i}(\boldsymbol{\mathit{l^{i+1}}}),
\end{aligned}
\end{equation}
where $\boldsymbol{j} \in \{0,1,2\}$, $\varepsilon_{i}$ represents the concatenation operation on the high-frequency components, $\eta_{i}$ is a \(3 \times 3\) convolutional layer designed to transform the concatenated part to the same dimension as the low-frequency components for subsequent operations. 
$F_{i}(\cdot)$, $H_{i}(\cdot)$ and $G_{i}(\cdot)$ are arbitrary functions, and $\sigma_{i}(\cdot)$ is the sigmoid function. Since video compression primarily targets the high-frequency signal of the video, the changes in the low-frequency signal are relatively minor. Based on this, the distortion layer simulates the compression of a low-frequency signal by continuously adding perturbations to the original low-frequency signal. For high-frequency signals, a more complex approach is used for simulation. After passing through the last $\text{DWT\_INB}$. The low-frequency signal $\boldsymbol{\mathit{l}^{n+1}}$ and the high-frequency signals $\boldsymbol{\mathit{h}^{n+1}}$ are output, and then an Inverse Discrete Wavelet Transform (IDWT) is applied to generate the video $\boldsymbol{\mathit{s_{\text{noise}}}}$ that has been simulated to undergo HEVC compression processing. Thus, the $s_{noise}$ can be obtained by
\begin{equation}
\boldsymbol{\mathit{s_{\text{noise}}}} = \text{IDWT}(\boldsymbol{\mathit{l^{n+1}}}, \boldsymbol{\mathit{h^{n+1}}}).
\end{equation}

During the training process of the distortion layer, it is necessary to closely imitate the compression of HEVC. Therefore, in the forward propagation process, the video distortion $\boldsymbol{\mathit{v_{\text{noise}}}}$ in the network output should be as similar as possible to the video $\boldsymbol{\mathit{y_{\text{com}}}}$ compressed by HEVC. Correspondingly, in the backward process, the restored video $\boldsymbol{\mathit{y_{\text{rev}}}}$ should be as close as possible to the original uncompressed video $\boldsymbol{\mathit{v_{\text{origin}}}}$. A loss function $\boldsymbol{\mathit{L_{\text{noise}}}}$ is defined to minimize the average error between each pair of videos in the training data, and its expression is as follows: 
\begin{equation}
\boldsymbol{\mathit{L_{\text{noise}}}} = \text{MSE}(\boldsymbol{\mathit{v_{\text{noise}}}}, \boldsymbol{\mathit{y_{\text{com}}}}) + \text{MSE}(\boldsymbol{\mathit{y_{\text{rev}}}}, \boldsymbol{\mathit{v_{\text{origin}}}}),
\end{equation}
where Mean Squared Error\text{(MSE)} measures the difference between $\boldsymbol{\mathit{v_{\text{noise}}}}$ and $\boldsymbol{\mathit{y_{\text{com}}}}$ or $\boldsymbol{\mathit{y_{\text{rev}}}}$ and $\boldsymbol{\mathit{v_{\text{origin}}}}$, $\boldsymbol{\mathit{v_{\text{noise}}}}$ is the distorted video obtained by passing the original video $\boldsymbol{\mathit{v_{\text{origin}}}}$ through the distortion layer, and $\boldsymbol{\mathit{y_{\text{rev}}}}$ is the restored video obtained by passing the compressed video $\boldsymbol{\mathit{y_{\text{com}}}}$ through the distortion layer in the backward process. The specific calculation formulas are as follows: 
\begin{equation}
\boldsymbol{\mathit{v_{\text{noise}}}} = \text{IDWT}(N^+(\text{DWT}(\boldsymbol{\mathit{v_{\text{origin}}}}))),
\end{equation}
\begin{equation}
 {\boldsymbol{\mathit{y_{\text{rev}}}} = \text{IDWT}(N^-(\text{DWT}(\boldsymbol{\mathit{y_{\text{com}}}}))),}
\end{equation}
where $DWT(\cdot)$ and $IDWT(\cdot)$ represent the Discrete Wavelet Transform and its inverse transform, respectively. $N^+(\cdot)$ and $N^-(\cdot)$ represent the forward encoding and backward decoding of the distortion layer. After the distortion layer is trained, to ensure the stability of the distortion layer and the convergence of the watermarking network, all parameters of the distortion layer need to be fixed. This allows the distortion layer to provide a stable distortion effect during the subsequent watermark embedding and extraction processes.

\subsection{Discriminator}
This paper employs a multiscale video discriminator from TGAN2\cite{tgan}, which consists of four 3D residual networks capable of capturing features of the watermarked video $\boldsymbol{\mathit{s}}$ from various perspectives. The working principle of the discriminator is as follows. First, the watermarked video $\boldsymbol{\mathit{s}}$ undergoes global pooling, which helps extract key information of the video across different space-time dimension. These pieces of information are then fed into four independent 3D residual networks, each of which focuses on capturing specific features of the video. Through this multi-scale analysis, the discriminator can more comprehensively understand and evaluate the characteristics of the watermarked video. Finally, features extracted from each 3D residual network are aggregated and passed to a fully connected layer for the final discrimination decision. The mathematical expression can be represented as:
\begin{equation}
x = \text{dis}(\boldsymbol{\mathit{s}}),
\end{equation}
where $\boldsymbol{\mathit{s}}$ represents the watermarked video, $dis(\cdot)$ represents the discriminator model, and ${\mathit{x}}$ represents the output of $\boldsymbol{\mathit{s}}$ after it passes through the discriminator. 

The empirical observation is that good temporal consistency in the video watermarking framework is crucially achieved using a powerful video discriminator. Under the video watermarking framework, the discriminator is tasked with not only identifying the presence of watermark but also enhancing the quality of the video. By employing the discriminator, the watermarking model can maintain video quality while effectively resisting compression, noise, and other potential attacks. This thereby enhances the security and reliability of the watermarking.

\subsection{Loss Function}

1) $\boldsymbol{\mathit{Video\ loss:}}$
The core task of the video forward encoding process is to embed the watermark $\boldsymbol{\mathit{M}}$ into the cover video $\boldsymbol{\mathit{c}}$, generating a watermarked video $\boldsymbol{\mathit{s}}$. To achieve the invisibility of the watermark, this paper defines a video loss function $\boldsymbol{\mathit{L_{\text{video}}}}$ to quantify the difference between the watermarked video and the cover video. The expression for the video loss function is as follows: 
\begin{equation}
\boldsymbol{\mathit{L_{\text{video}}}} = \text{MSE}(\boldsymbol{\mathit{c}}, \boldsymbol{\mathit{s}}).
\end{equation}

2) $\boldsymbol{\mathit{Message\ loss:}}$
In the backward decoding stage, to extract the watermark $\boldsymbol{\mathit{M}}$ from the distorted video $\boldsymbol{\mathit{s_{\text{noise}}}}$, a message loss function $\boldsymbol{\mathit{L_{\text{message}}}}$ is defined, and the expression of which is as follows:
\begin{equation}
\boldsymbol{\mathit{L_{\text{message}}}} = \text{MSE}(\boldsymbol{\mathit{M_{\text{ex}}}}, \boldsymbol{\mathit{M}}).
\end{equation}

3) $\boldsymbol{\mathit{Discriminator\ loss:}}$
To enhance the imperceptibility of the watermark and improve the video quality, a more complex discriminator is introduced to achieve these goals. The loss function $\boldsymbol{\mathit{L_{\text{dis}}}}$ is defined as follows:
\begin{equation}
\boldsymbol{\mathit{L_{\text{dis}}}} = \text{CrossEntropy}(\text{dis}(\boldsymbol{\mathit{s}}), \text{label}),
\end{equation}
where $\text{label}$ represents the ground truth labels indicating whether the videos are generated video or real video.

4) $\boldsymbol{\mathit{Total\ loss:}}$
The total training loss can be ﬁnally deﬁned as the weighted sum of video loss $\boldsymbol{\mathit{L_{\text{video}}}}$, message loss $\boldsymbol{\mathit{L_{\text{message}}}}$ and discriminator loss $\boldsymbol{\mathit{L_{\text{dis}}}}$.
\begin{equation}
\label{total_loss}
\boldsymbol{\mathit{L_{\text{total}}}} = \lambda_1 \boldsymbol{\mathit{L_{\text{video}}}} + \lambda_2 \boldsymbol{\mathit{L_{\text{message}}}} + \lambda_3 \boldsymbol{\mathit{L_{\text{dis}}}},
\end{equation}
where $\lambda_1$,$\lambda_2$ and $\lambda_3$ are weight parameters used to balance the contributions of various loss components. {Specifically, \( \lambda_1 \) controls the video quality, \( \lambda_2 \) regulates the watermark embedding strength, and \( \lambda_3 \) is used for adversarial training.} To avoid instability in model performance during the initial training phase, a staged training strategy is adopted in this paper. 
During the pre-training phase, the focus is on minimizing the video loss $\boldsymbol{\mathit{L_{\text{video}}}}$ and the message loss $\boldsymbol{\mathit{L_{\text{message}}}}$, while the adversarial loss
$\boldsymbol{\mathit{L_{\text{dis}}}}$ is not considered, $\mathit{i.e.}$, the weight $\lambda_3$ is set to 0. This strategy is designed to prevent the model collapse that might be caused by introducing adversarial training too early, avoiding unnecessary complexity. By employing this staged training approach, the model is ensured to learn stably at each training stage and gradually improve its performance, ultimately achieving optimized training results.

\section{EXPERIMENTS}
{The experimental dataset used is UCF-101, which is a large-scale video dataset containing 101 action categories with a total of 13,320 video samples. {Considering the presence of a large number of video with similar scenes in the dataset, this paper carefully selected 2,000 videos with low scene similarity for model training, additionally, we selected 500 videos for evaluation.} In order to enhance the generalization ability of the model in diverse scenarios and training efficiency. At the same time, to comprehensively evaluate performance of the model, the Kinetics 600 and REDS datasets are employed in this paper to assess model identification and processing capabilities of the new data. Kinetics-600 is a large-scale dataset containing diverse human actions in real-world settings, making it one of the most widely used datasets in the field. The REDS dataset, on the other hand, is specifically designed for video super-resolution and contains high-resolution videos with detailed textures and complex movements.} All of them can be downloaded from \url{https://www.crcv.ucf.edu/data/UCF101.php}, \url{https://seungjunnah.github.io/Datasets/reds} and \url{https://deepmind.google/kinetics}. 

{The training process can be divided into two distinct stages. In the first stage, no discriminator is introduced. The hyperparameters are set as follows: \( \lambda_1 = 1.0 \), \( \lambda_2 = 10.0 \), and \( \lambda_3 = 0.0 \). In the second stage, the hyperparameters are adjusted to \( \lambda_1 = 1.0 \), \( \lambda_2 = 2.0 \), and \( \lambda_3 = 0.0001 \). For both stages, the learning rate is set to 0.0001, and the Adam optimizer is employed with \( \beta_1 = 0.5 \) and \( \beta_2 = 0.999 \) to facilitate convergence. It should be noted that in the second stage, an exponential decay with a rate of 0.5 is applied every 20 epochs, and this experiment is implemented by Pytorch and trained on an NVIDIA Tesla V100 GPU. }

During the training phase, to enhance the adaptability of the model to different video sizes, this paper implements a random cropping strategy. Specifically, after cropping, the video clips are resized to a dimension of \(3 \times 8 \times 128 \times 128\), where 3 represents the number of channel dimensions, 8 represents the number of video frames, and \(128 \times 128\) is the spatial resolution of each frame. This random cropping scheme not only increases the diversity of the training data but also simulates the various video sizes that may be encountered in practical applications, thereby improving the generalization ability of the model. 

\subsection{Comparison to Existing Methods}
{This paper employs metrics such as Peak Signal-to-Noise Ratio (PSNR), Learned Perceptual Image Patch Similarity (LPIPS), temporal location precision (tLP) and Accuracy (ACC) to evaluate the performance of different watermarking schemes in terms of video quality, temporal consistency and the accuracy of watermark extraction. Larger PSNR imply better video quality, lower LPIPS and tLP\cite{tlp} values indicate better visual quality and temporal consistency,} and higher ACC values suggest that the model can extract the embedded watermark with greater accuracy. Additionally, due to the structure of the INN network, DINVMark is more suited to embedding watermark in the form of regular rectangular shapes, such as \(8 \times 8\) , \(16 \times 16\) and \(32 \times 32\) square watermarks. For irregular rectangular watermark, such as \(8 \times 12\) , \(8 \times 14\) and \(8 \times 16\) rectangular watermarks, DINVMark needs to add linear layers before and after the network to map the watermark to a square watermark template before embedding. This lead to insufficient coupling between the encoder and decoder, resulting in better model performance when embedding regular rectangular watermarks compared to irregular rectangular watermarks. Therefore, in the subsequent experimental section, this paper will conduct experimental analysis on the two forms of watermark mentioned above.

\begin{figure*}[!t]
\centering
\subfloat[\textrm{Original}]{\includegraphics[width=2.2in]{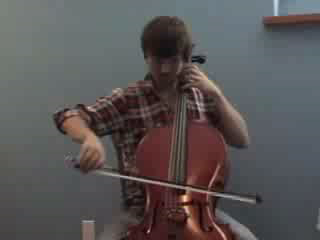}
\label{origin}}
\hfil
\subfloat[\textrm{Gaussian Noise(std=0.04)}]{\includegraphics[width=2.2in]{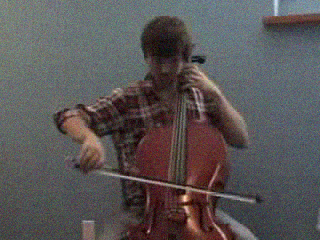}%
\label{gn}}
\hfil
\subfloat[\textrm{HEVC(QP=22)}]{\includegraphics[width=2.2in]{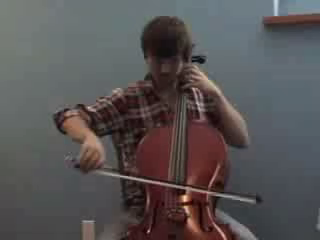}
\label{hevc}}
\caption{Sample frames from distorted videos}
\label{noiseImg}
\end{figure*}

\begin{table*}
\centering
\caption{{Comparison of ACC(\%) under different distortion attacks}}
\label{comparison}
\begin{tabular}{lccccccc}
\toprule
Dataset & Framework & \makecell{Frame Average \\ ($N$ = 3)} & \makecell{Frame Drop \\ ($p$ = 0.5)} & \makecell{Frame Swap \\ ($p$ = 0.5)} & \makecell{Gaussian Noise \\ (std = 0.04)} & \makecell{H.264/AVC \\ (CRF = 22)} & \makecell{HEVC \\ (QP = 22)} \\
\midrule
{Kinetics 600} & {HiDDeN[14]} & {81.02} & {78.61} & {81.76} & {82.23} & {65.78} & {69.14} \\
 & {DT-CWT[8]} & {95.80} & {97.70} & {98.70} & {95.65} & {\underline{96.91}} & {97.74} \\
 & {DVMark[19]} & {98.10} & {98.99} & {99.35} & {98.56} & {92.94} & {-} \\
 & {REVMark[20]} & {\underline{99.98}} & {\underline{99.81}} & {\underline{99.98}} & {\underline{99.97}} & {95.84} & {\underline{98.15}} \\
 & {DINVMark(ours)} & {\textbf{100.00}} & {\textbf{100.00}} & {\textbf{100.00}} & {\textbf{100.00}} & {\textbf{98.66}} & {\textbf{99.94}} \\
\midrule
UCF-101 & HiDDeN[14] & 81.53 & 78.61 & 82.72 & 82.91 & 64.10 & 68.76 \\
 & REVMark[20] & \underline{99.99} & \underline{99.99} & \underline{99.70} & \underline{99.50} & \underline{95.66} & \underline{98.34} \\
 & DINVMark(ours) & \textbf{100.00} & \textbf{100.00} & \textbf{100.00} & \textbf{100.00} & \textbf{98.45} & \textbf{99.98} \\
\midrule
REDS & HiDDeN[14] & 79.14 & 78.35 & 82.71 & 85.06 & 64.17 & 69.85 \\
 & REVMark[20] & \underline{99.96} & \underline{99.41} & \underline{99.96} & \underline{99.61} & \underline{95.29} & \underline{98.21} \\
 & DINVMark(ours) & \textbf{100.00} & \textbf{100.00} & \textbf{100.00} & \textbf{100.00} & \textbf{97.54} & \textbf{100.00} \\
\bottomrule
\end{tabular}
\end{table*}

$\textbf{1) Robustness Evaluation:}$
To assess the robustness of the watermark, this paper tests the accuracy of watermark extraction under various attacks, including frame averaging, frame dropping, frame swapping, Gaussian noise, H.264/AVC compression, and HEVC compression. The experimental results are shown in Table \ref{comparison}, where frame averaging involves averaging all frames within a temporal window, with $N$ referring to the number of frames in the window. In frame dropping and frame swapping, $p$ denotes the probability of a frame being dropped (or swapped with its immediate neighbor). The $std$ in Gaussian noise refers to the standard deviation of the Gaussian kernel. In H.264/AVC compression and HEVC compression, QP stands for the Quantizer Parameter. All results in this section are evaluated based on video sizes of \(3 \times 8 \times 128 \times 128\). Fig. \ref{noiseImg} illustrates the video frames after different attacks.

{Table \ref{comparison} presents the accuracy of watermark for the HiDDeN, DT-CWT, DVMark, REVMark and DINVMark schemes after embedding 96-bit watermark and under various distortion attacks, where boldface indicates the best results, and underline indicates the second-best results. As can be seen from Table \ref{comparison}, DINVMark outperforms the comparative schemes in terms of watermark accuracy across three datasets, and in most cases, it achieves 100\% accuracy even when facing different distortion attacks. This demonstrates the excellent robustness of DINVMark. To further verify the generalization capability of DINVMark across different datasets, we conducted experiments using the classic HiDDeN and the state-of-the-art REVMark. The results indicate that DINVMark also exhibits superior performance across various datasets. By comparing performance of our model across different datasets, it can be observed that our model achieves the best performance on the REDS dataset after compression using HEVC, but is not optimal when compressed with H.264/AVC. This discrepancy is attributed to the characteristics of the REDS dataset, which consists of high-resolution videos with detailed textures and complex movements. Such videos are more susceptible to losing watermarks under higher compression intensity, while they are more likely to preserve watermarks under weaker compression intensity. Through analysis, we find that the superior performance of DINVMark is primarily attributed to the unique architecture of the INN and distortion layer design. In contrast, REVMark achieves such performance by introducing the TAsBlock before feature extraction, which enables more effective capture of inter-frame differences. while DVMark introduces a novel multiscale design in both the encoder and decoder. This design allows messages to be embedded across multiple spatial-temporal scales, thereby achieving higher accuracy and robustness.}

\begin{figure*}[!t]
    \centering
    \subfloat[]{
        \includegraphics[width=0.46\textwidth]{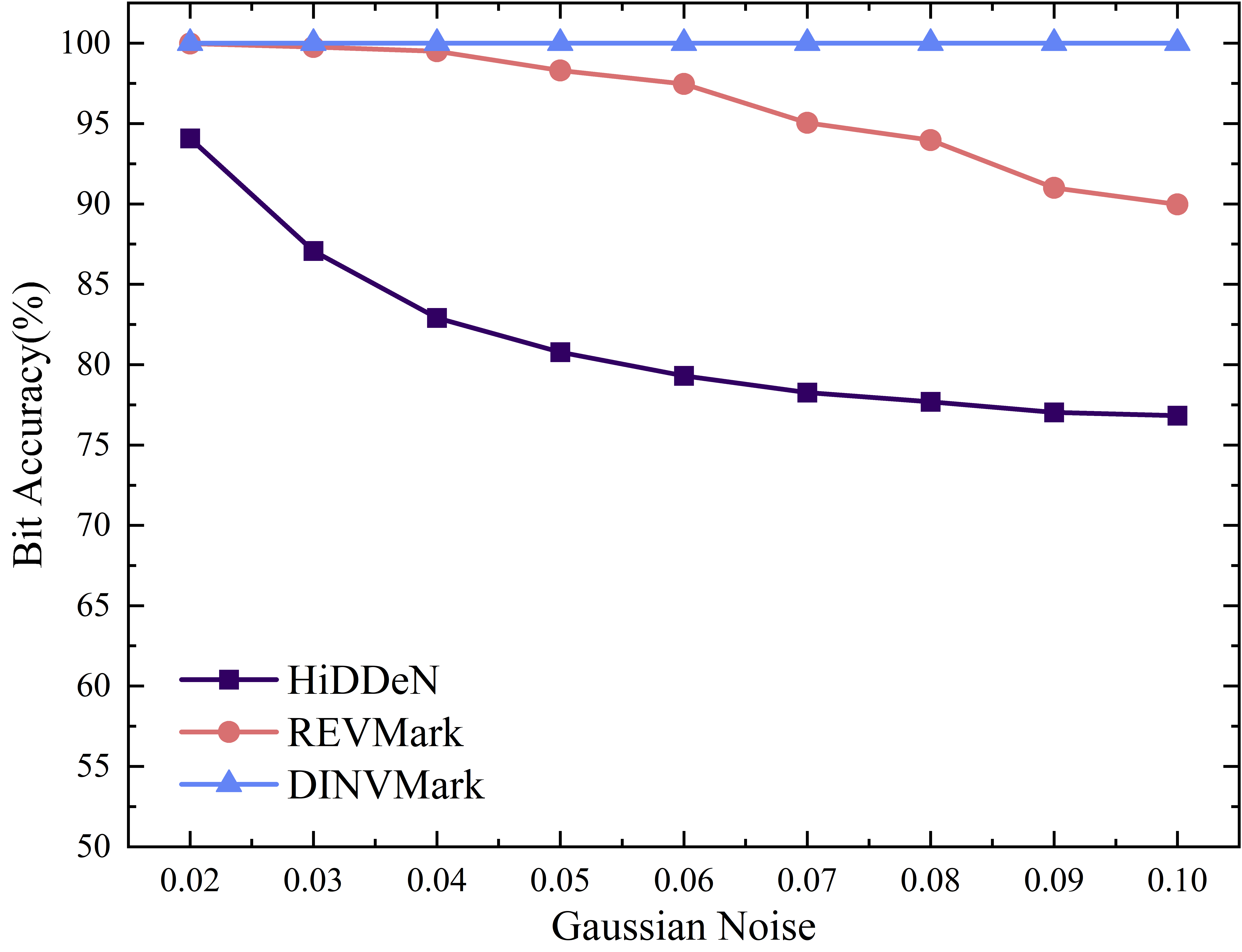}
        \label{gn_compare}
    }
    \hfill
    \subfloat[]{
        \includegraphics[width=0.46\textwidth]{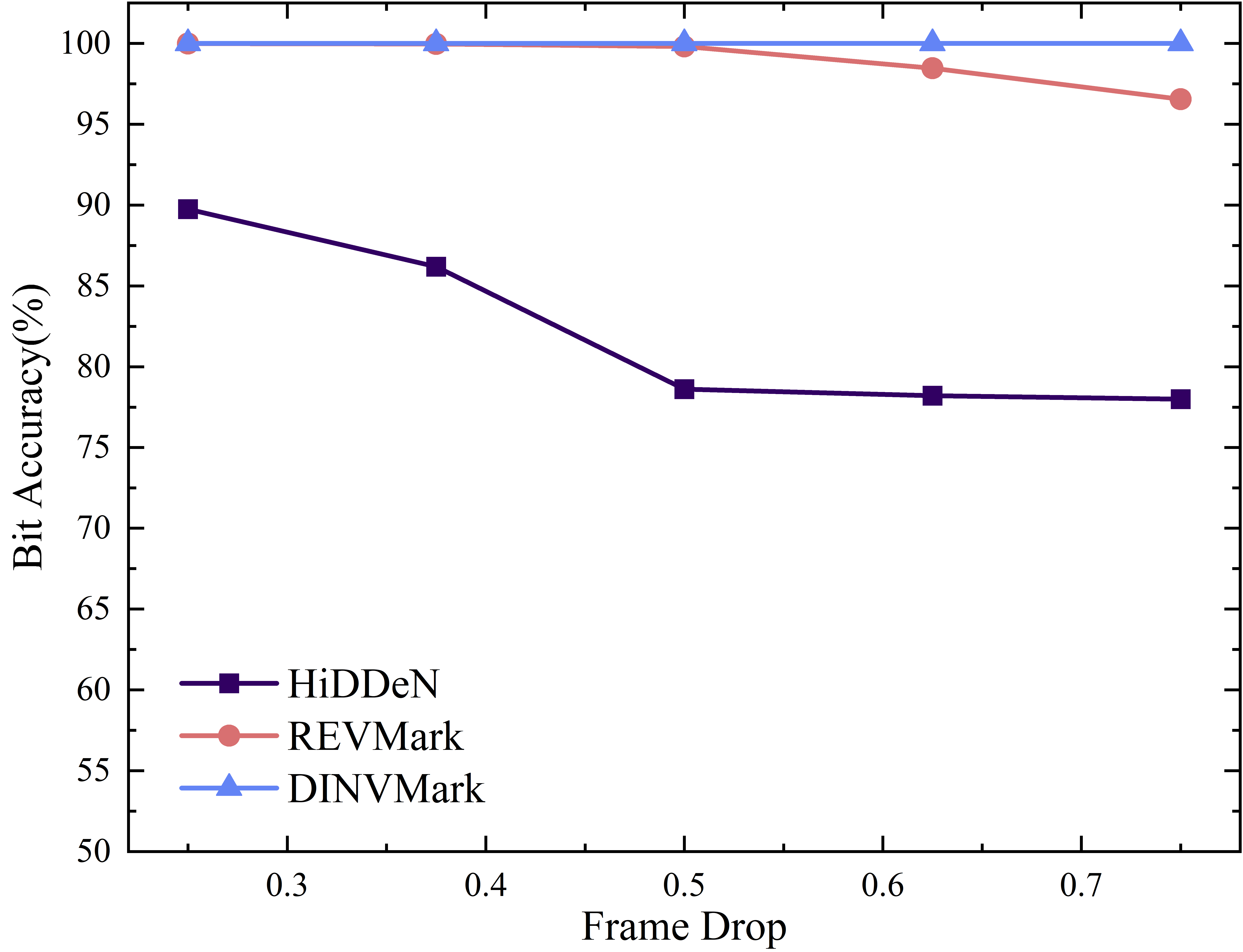}
        \label{drop}
    }
    \vspace{0.3cm}
    \subfloat[]{
        \includegraphics[width=0.46\textwidth]{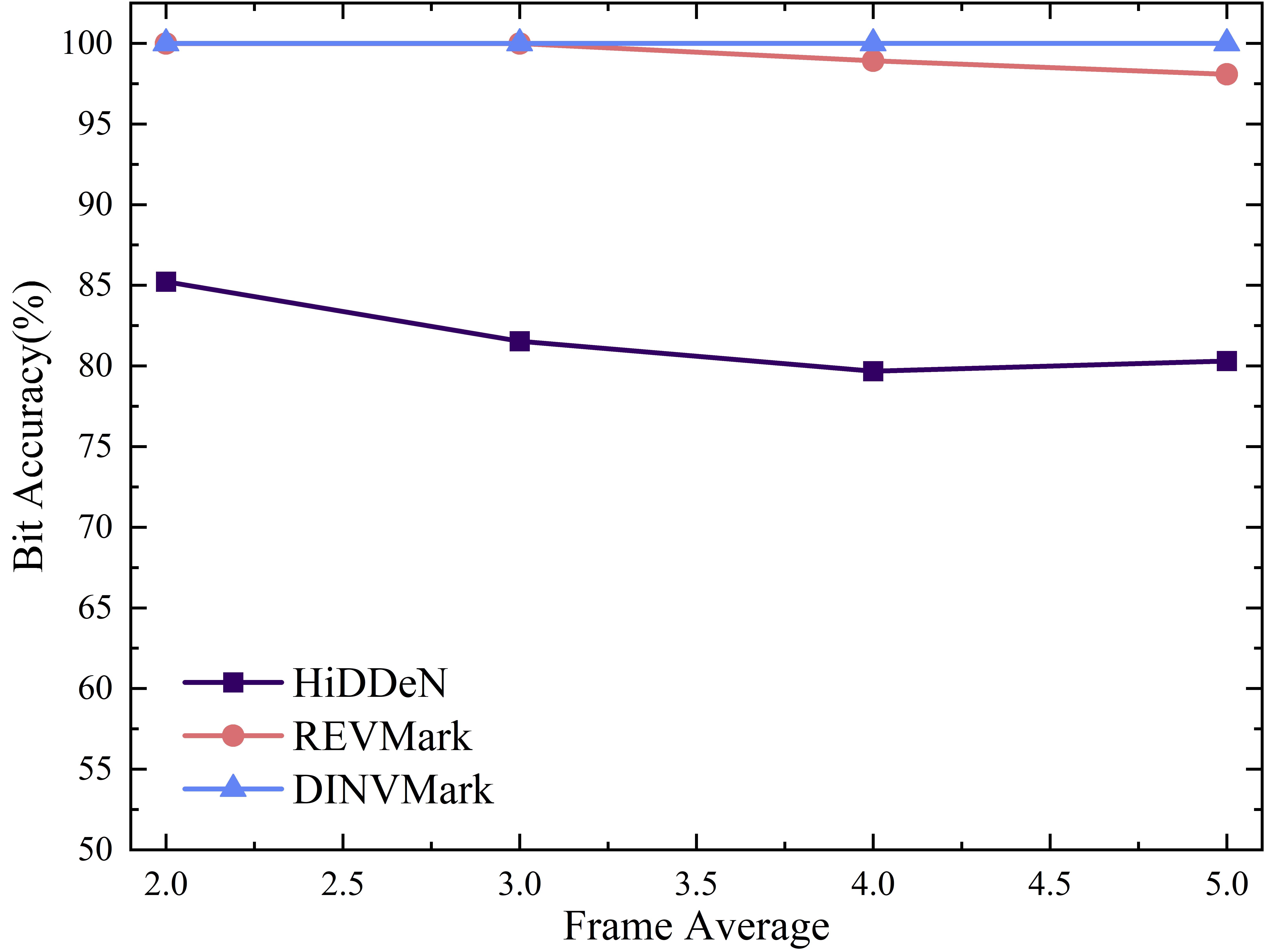}
        \label{average}
    }
    \hfill
    \subfloat[]{
        \includegraphics[width=0.46\textwidth]{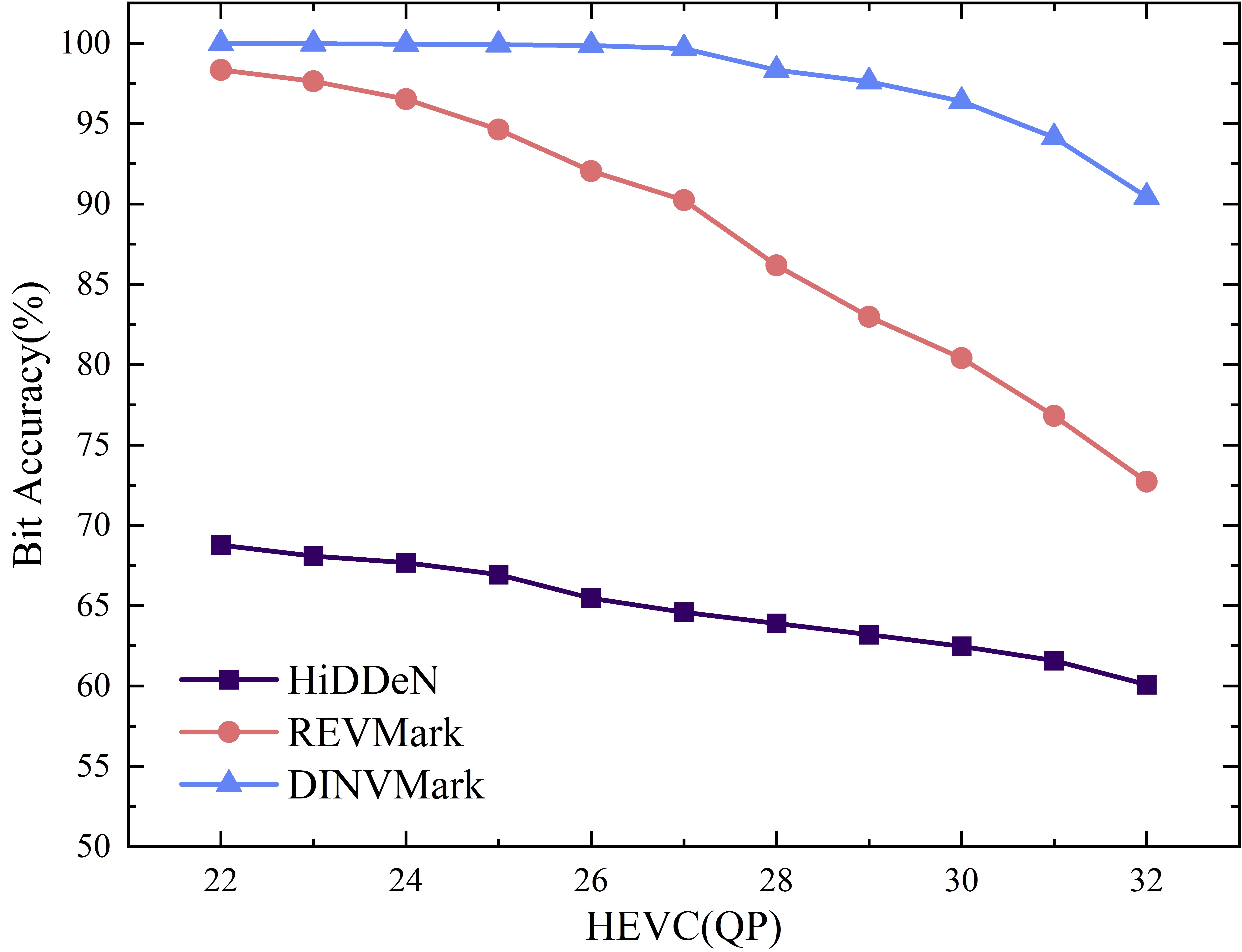}
        \label{qphevc}
    }
    \caption{{Comparing the bit accuracy of different watermarking schemes under varying degrees of distortion attacks}}
    \label{com_many}
\end{figure*}

{To further validate the robustness of DINVMark against distortion attacks such as HEVC compression, this paper simulates the watermark extraction accuracy of various schemes when embedding a 96-bit watermark. The final experimental results are shown in Fig. \ref{com_many}. As can be seen from Fig. \ref{com_many}, even under high-intensity attack conditions, DINVMark can still maintain high accuracy. Notably under high compression conditions where QP is increased to 32, DINVMark can still maintain a 90\% accuracy rate.} This is mainly attributed to the design of the distortion layer and the network structure of the INN. Firstly, the distortion layer effectively enhances the robustness of the embedded watermark by simulating the HEVC compression process, enabling it to better resist the interference caused by compression. Secondly, the INN network structure allows the encoder and decoder to share the same parameters, enabling the decoder to more accurately restore the watermark embedded by the encoder, thereby improving the accuracy of watermark extraction.

\begin{table}[htbp]
\centering
\caption{{ACC(\%) for different compression strengths for a range of watermark lengths}}
\label{watermarkLength}
\scalebox{0.8}{
\begin{tabular}{cccccccc}
    \toprule
    \multirow{2}{*}{QP} & \multicolumn{3}{c}{Irregular rectangular watermark} & & \multicolumn{3}{c}{Regular rectangular watermark} \\
    \cmidrule(r){2-4} \cmidrule(l){6-8}
    & 96 & 112 & 128 & & 64 & 256 & 1024 \\
    \midrule
    22 & {99.98} & {99.99} & {99.90} & & {99.10} & {95.63} & {94.06} \\
    27 & {99.74} & {99.85} & {99.37} & & {97.84} & {94.09} & {93.32} \\
    32 & {90.44} & {91.04} & {92.48} & & {91.92} & {89.47} & {68.65} \\
    \bottomrule
\end{tabular}
}
\end{table}

To enhance the robustness of the watermarking model against varying degrees of HEVC compression, the distortion layer is trained using videos compressed to different extents, allowing it to learn the patterns of video compression at different levels. Building on this, the model is further trained end-to-end to optimize the extraction of the watermark under various compression conditions. From Table \ref{watermarkLength}, it can be observed that when embedding irregular rectangular watermark, the accuracy is similar under different levels of video compression due to the similar watermark embedding capacity. When embedding regular rectangular watermark, it can be observed that even after embedding 256-bit or even 1024-bit watermarks and subsequent HEVC compression processing, this method is still capable of effectively extracting watermark while ensuring video quality. Nevertheless, when QP reaches 32 and the watermark embedding capacity increases to 1024 bits, the accuracy significantly decreases, because embedding larger capacity watermark requires more modifications to the video, and combined with the increased compression intensity, it leads to a significant decrease in the quality of the watermarked video, resulting in the loss of watermark. In such cases, the accuracy of the watermark drops significantly, indicating that the watermark capacity and compression intensity have a considerable impact on the robustness of watermark extraction.

\begin{table}[htbp]
\centering
\caption{{Quality comparison of the watermarked videos using different watermark schemes}}
\label{quality}
\begin{tabular}{lccc}
\toprule
Framework & PSNR(dB)$\uparrow$ & LPIPS$\downarrow$ & tLP$\downarrow$\\
\midrule
HiDDeN[14] & {34.45} & {0.1448} & {0.0254}\\
{DT-CWT[8]} & {36.91} & {0.0814} & {0.0083}\\
{DVMark[19]} & {37.00} & {0.0570} & {\underline{0.0016}}\\
REVMark[20] & {\underline{37.41}} & {\underline{0.0273}} & {0.0086}\\
DINVMark(ours) & {\textbf{37.50}} & {\textbf{0.0241}} & {\textbf{0.0014}}\\
\bottomrule
\end{tabular}
\end{table}

$\textbf{2) Evaluation of Visual Quality:}$
Table \ref{quality} presents the quality of watermarked videos for different schemes under the condition of embedding a 96-bit watermark.

From Table \ref{quality}, {it can be observed that, under the condition of embedding the same watermark capacity, DINVMark outperforms the comparison scheme in all metrics. This result is primarily attributed to the construction of the loss function and preprocessing of the video dataset. In contrast, REVMark achieves such performance by formulating the mask loss for deep video watermarking through the joint exploitation of spatial and temporal masks.}

\begin{figure*}[!t]
\centering
\subfloat[\textrm{Original}]{\includegraphics[width=0.24\textwidth]{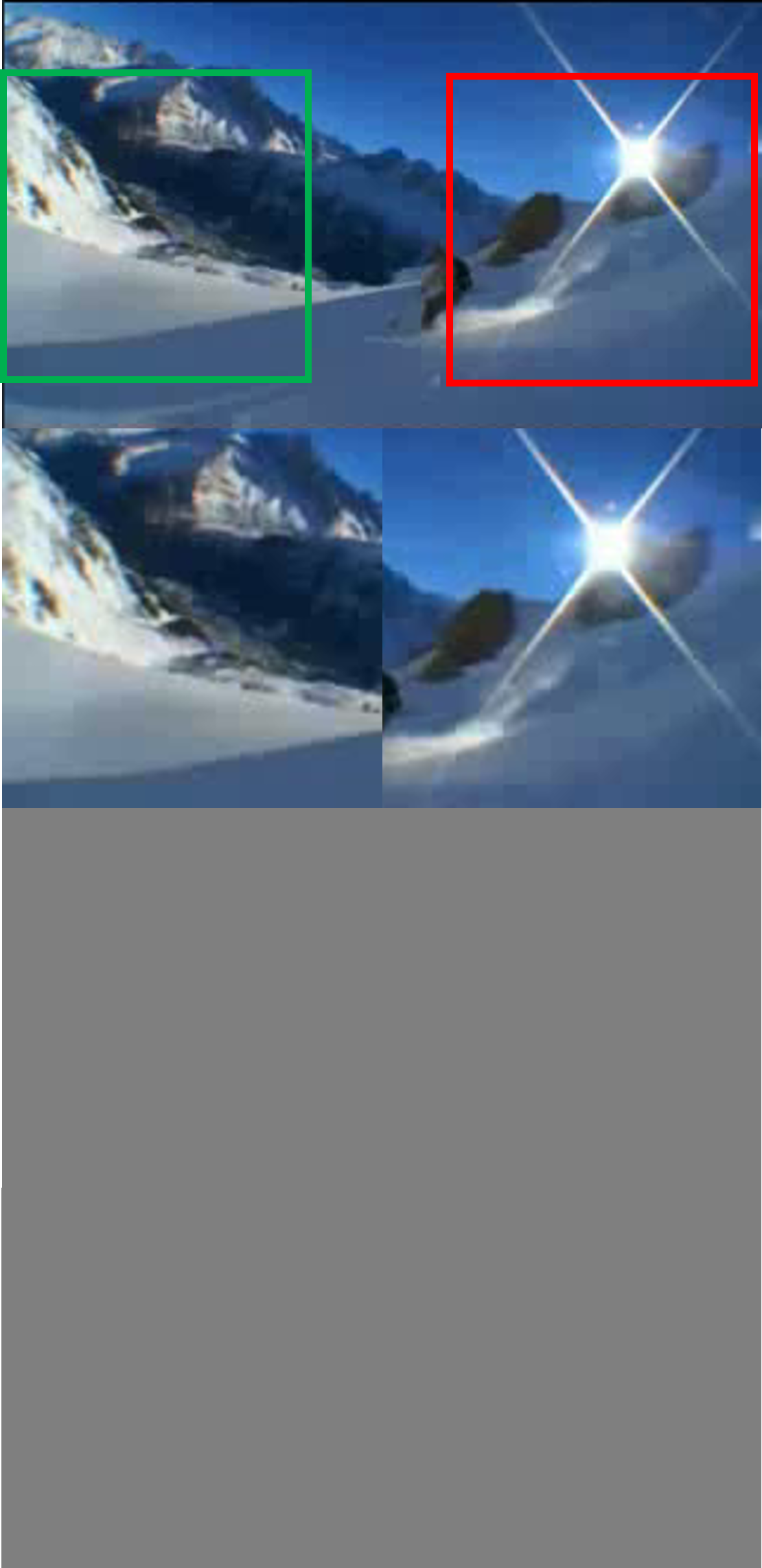}%
\label{original}}
\hfil
\subfloat[\textrm{HiDDeN}]{\includegraphics[width=0.24\textwidth]{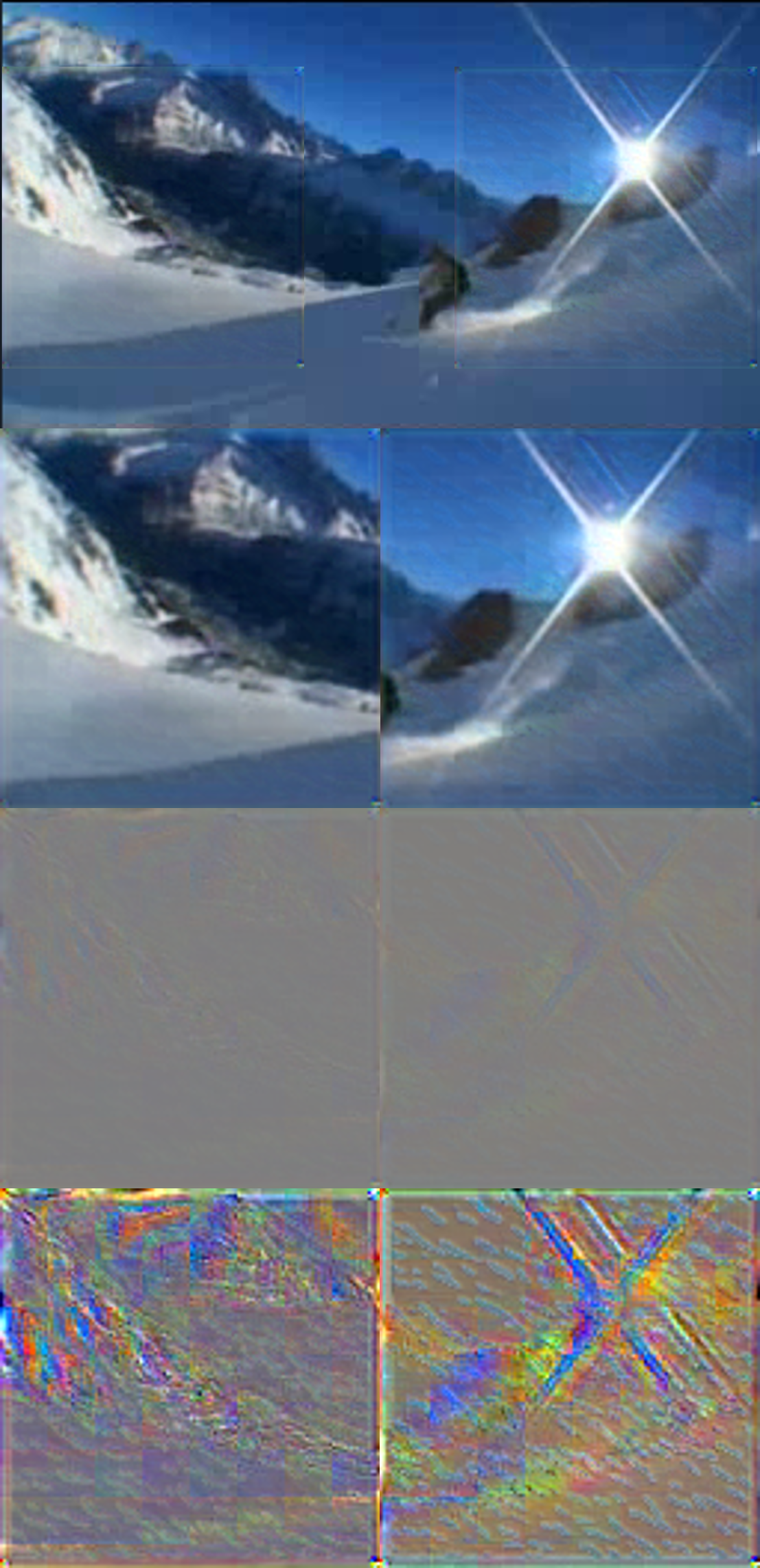}%
\label{hidden}}
\hfil
\subfloat[\textrm{REVMark}]{\includegraphics[width=0.24\textwidth]{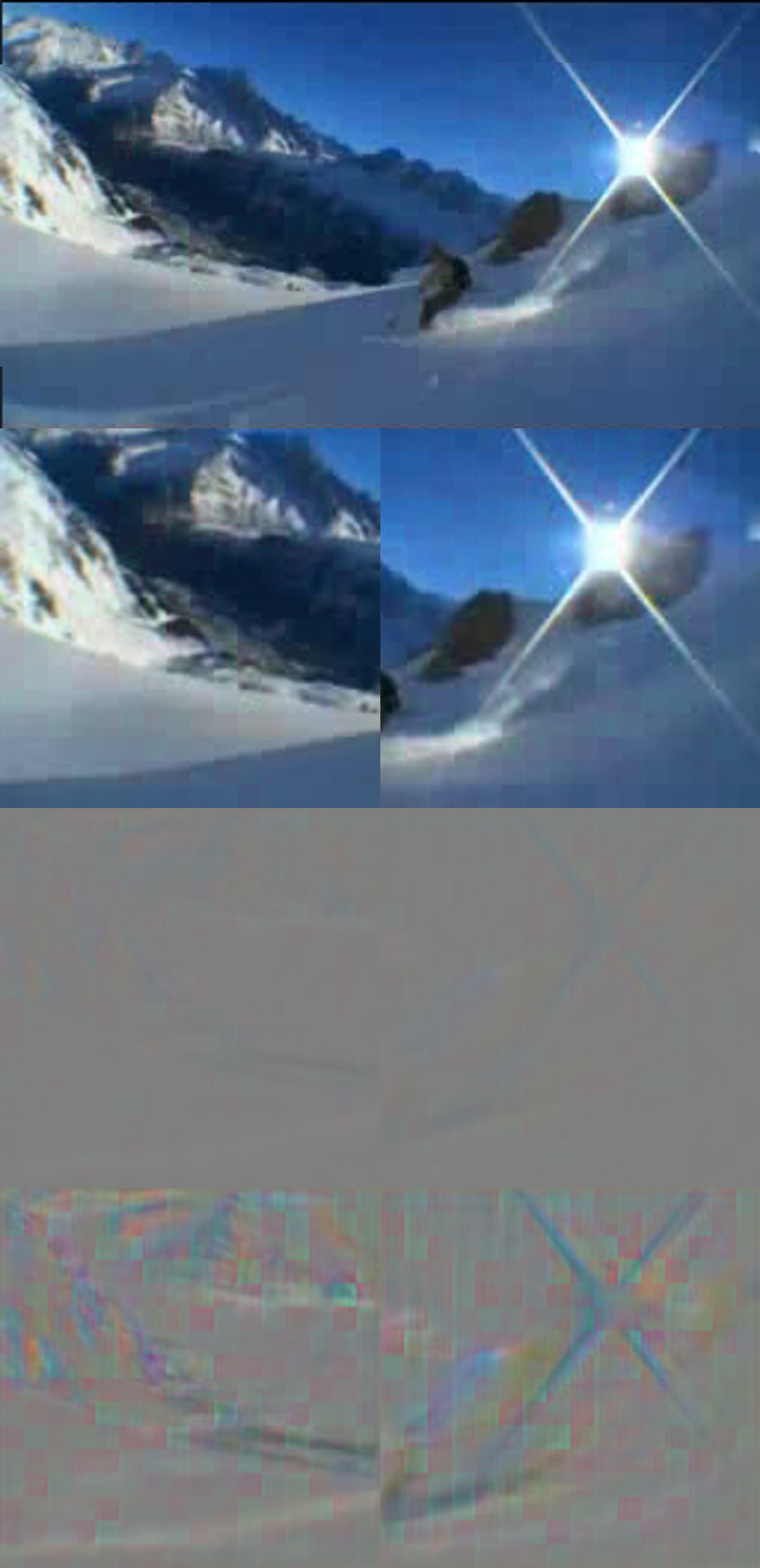}%
\label{revmark}}
\hfil
\subfloat[\textrm{DINVMark}]{\includegraphics[width=0.24\textwidth]{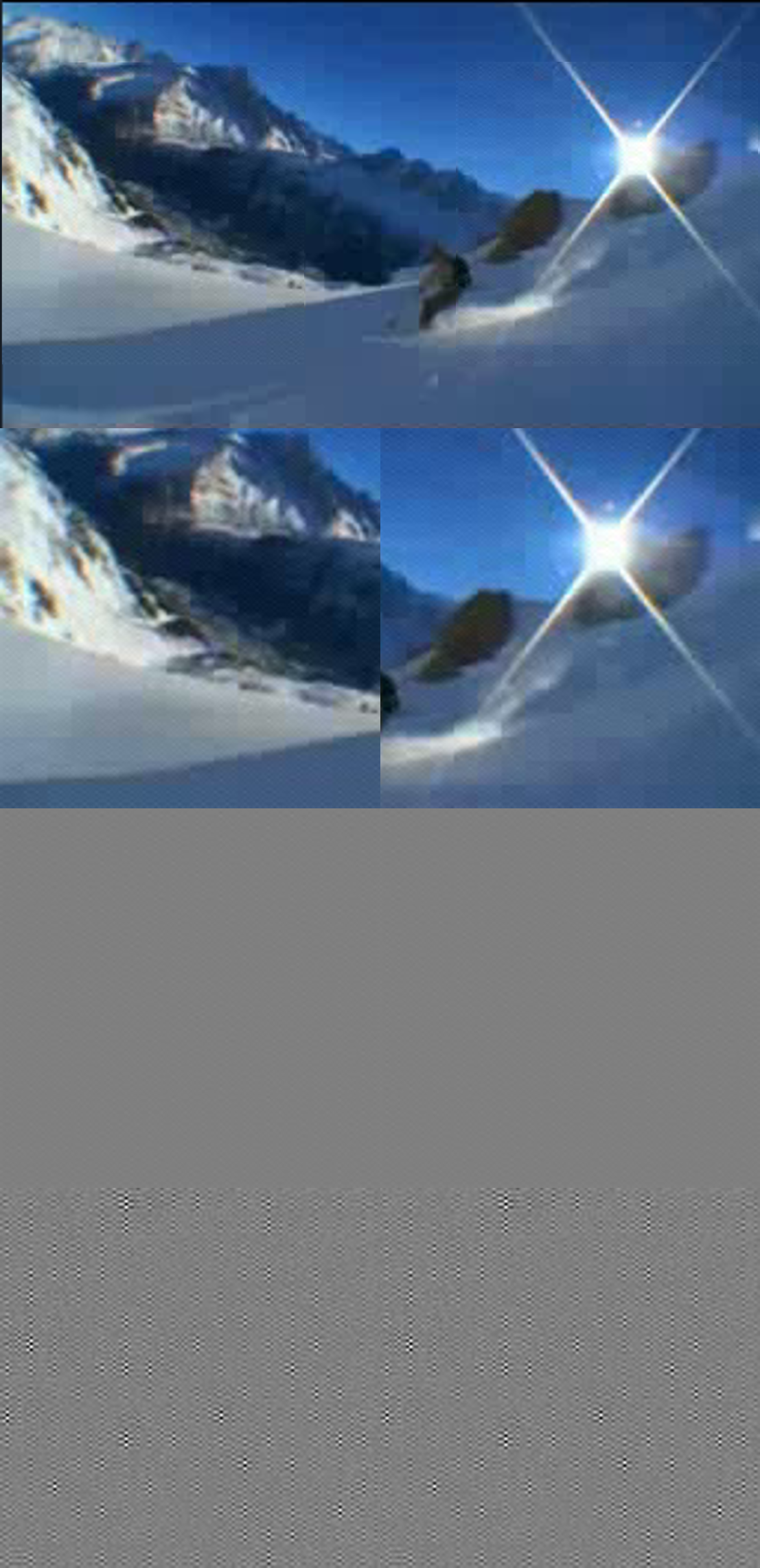}%
\label{mine}}
\caption{Qualitative visualization presentation for different watermark schemes}
\label{mergediff}
\end{figure*}


Fig. \ref{mergediff} presents the visualization of watermarked video frames for the HiDDeN, RevMark and DINVMark schemes, along with the difference between the watermarked video frames and the original video frames. Specifically, top row contains the frames
in full-view, second row contains two zoomed-in patches for each image, third row plots 
the normalized difference of the watermarked frames with the original frames and the bottom row amplify the different by 10 times. From Fig. \ref{mergediff}, it is evident that DINVMark exhibits superior visual quality. Even when the difference is magnified by 10 times, there is no significant blocky distortion or severe texture interference, indicating that the scheme causes minimal disruption to the original video. The low level of distortion is mainly attributed to the network structure of the INN, which better balances the capacity of the watermark and the visual quality of the video during the watermark embedding process. DINVMark can ensure the robustness of the watermark while minimizing visual disturbance, thus ensuring high visual quality of the video.

\subsection{Robustness-Quality-Payload Trade-off}
Comprehensive evaluation of watermark embedding networks requires assessment of model payload, robustness and quality metrics. This paper addresses the trade-offs between payload and video quality, and explores the interrelationships between these parameters and robustness. To assess robustness, we adopts ACC as the metric, while video quality is quantified through PSNR and LPIPS, and message length is directly related to payload. 

\begin{table}[htbp]
\centering
\caption{{PSNR (dB) versus ACC (\%) of DINVMark under HEVC Compression with QP=32}}
\label{psnr}
\scalebox{0.8}{
\begin{tabular}{lccccccc}
\toprule
PSNR (dB) & {38.87} & {38.40} & {37.86} & {37.50} & {37.20} & {36.71} & {36.20} \\
\midrule
ACC & {81.83} & {83.78} & {86.28} & {90.44} & {91.05} & {92.50} & {94.21} \\
LPIPS & {0.0182} & {0.0204} & {0.0234} & {0.0241} & {0.0265} & {0.0300} & {0.0438}\\
\bottomrule
\end{tabular}
}
\end{table}

$\textbf{1) Robustness-Quality Trade-off:}$
To explore the relationship between robustness and video quality, this paper fixed the payload at 96 bits and adjusted the watermark embedding strength $\lambda_2$ in Eq. \ref{total_loss} to observe its impact on video quality. As shown in Table \ref{psnr}, it demonstrates the correlation between ACC and PSNR under different watermark embedding strengths. Since the ACC of the model is stable at around 99\% for QP=22 and QP=27, QP=32 is chosen for this experiment. From Table \ref{psnr}, it is evident that as the watermark embedding strength increases, the robustness of the watermark gradually improves, while the visual quality of the video decreases. Because the model needs to embed more watermark into the video to enhance its resistance to attacks and extraction accuracy, which leads to a decline in video visual quality. This phenomenon conforms to the general rule of watermark embedding and also proves the effectiveness of the model.

\begin{figure*}[!t]
\centering
\subfloat[\textrm{Irregular rectangular watermark}]{\includegraphics[width=3in]{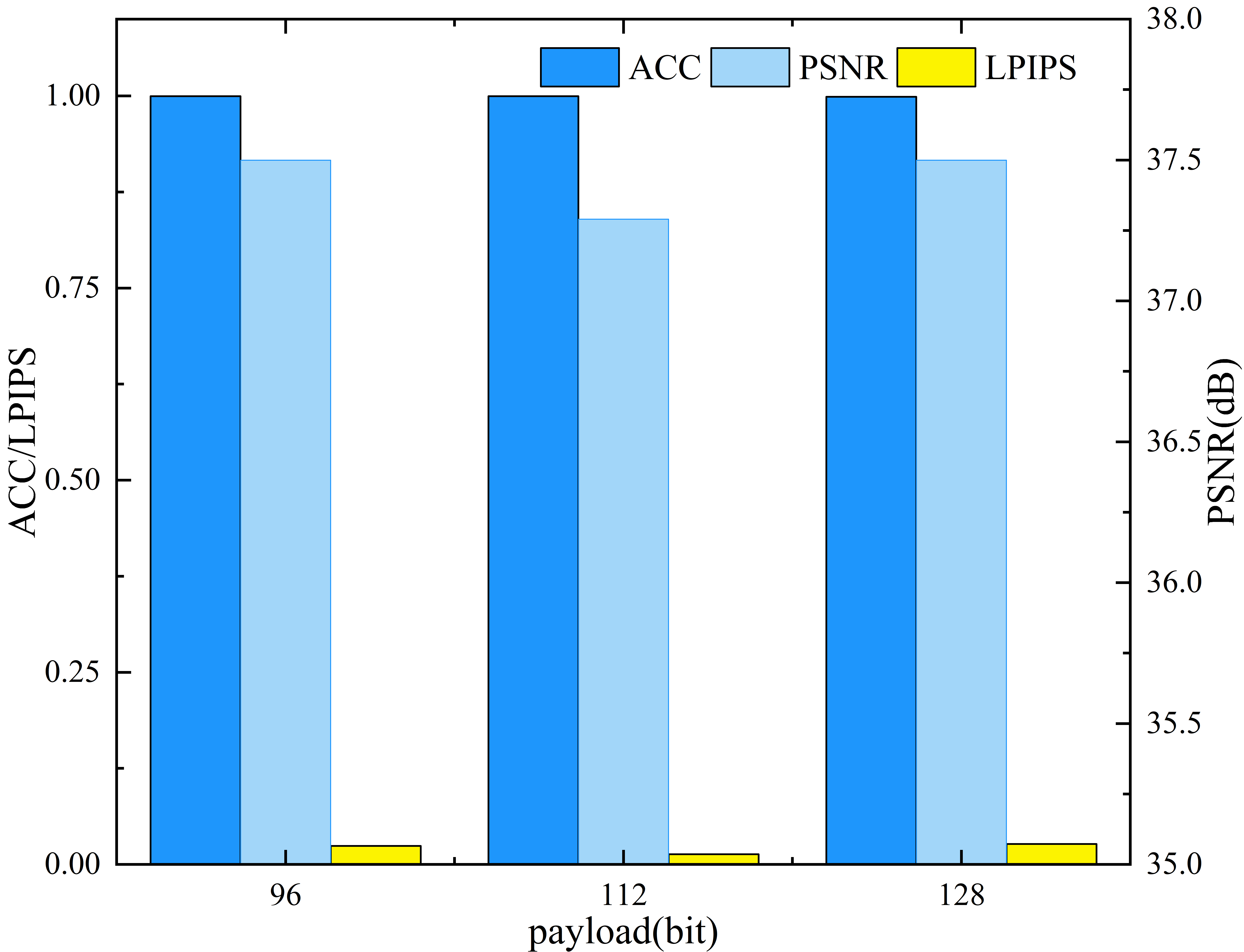}%
\label{payload_diff}}
\hfil
\subfloat[\textrm{Regular rectangular watermark}]{\includegraphics[width=3in]{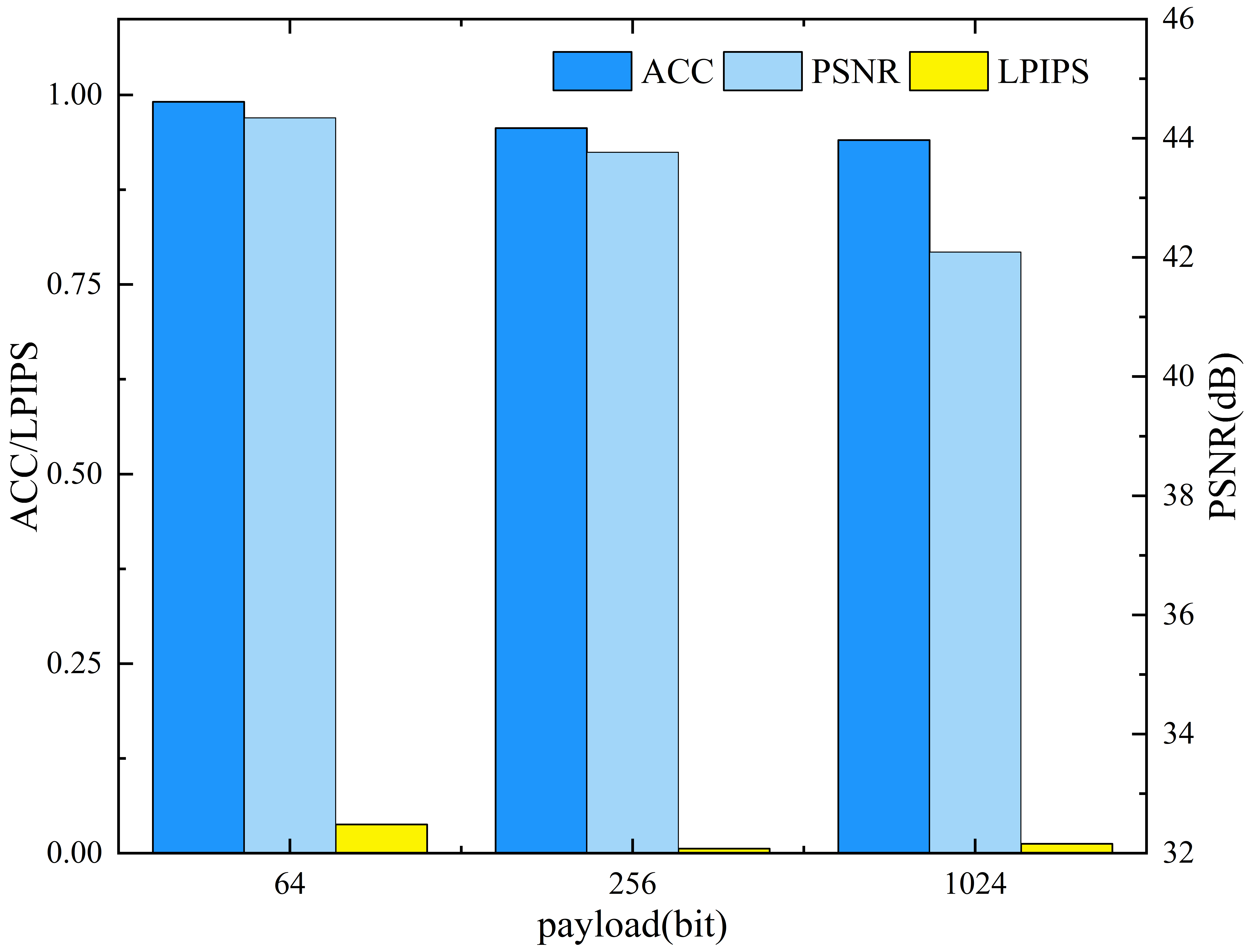}%
\label{payload}}
\caption{{The impact of embedding watermark capacity on video quality and accuracy when the QP is 22}}
\label{merge}
\end{figure*}

$\textbf{2) Robustness-Payload Trade-off:}$
This paper also explores the relationship between robustness and payload. The model is trained for different payloads and its ACC is evaluated under HEVC compression with QP =22. Furthermore, this section discusses the impact of watermark template shapes on the performance of DINVMark. In DINVMark, regular rectangular watermark input templates, such as \(8 \times 8\), \(16 \times 16\), and \(32 \times 32\), can be directly applied to the model. As for irregular rectangular watermark input templates, such as \(8 \times 12\), \(8 \times 14\), and \(8 \times 16\), a linear layer needs to be added before and after the network to complete the shape transformation of the watermark template. The results of the relationship between robustness and payload for DINVMark are shown in Fig. \ref{merge}.

From Fig. \ref{merge}, it can be seen that even when the payload reaches 1024 bits, the model still demonstrates good robustness, with PSNR maintained above 40 dB. Additionally, comparing Fig. \ref{merge}, it is observed that the model without linear layers performs better than the model with linear layers. The reason for this phenomenon is that the linear layers added at both ends of the model do not share parameters, which can lead to insufficient coupling: the larger the embedding capacity, the higher the evaluation metrics.

\subsection{Ablation Study}
To further verify the effectiveness of the proposed network, this experiment conducted an ablation study on the presence of the distortion layer and the number of INBs.

\begin{figure*}[!t]
\centering
\subfloat[\textrm{Irregular rectangular watermark}]{\includegraphics[width=3in]{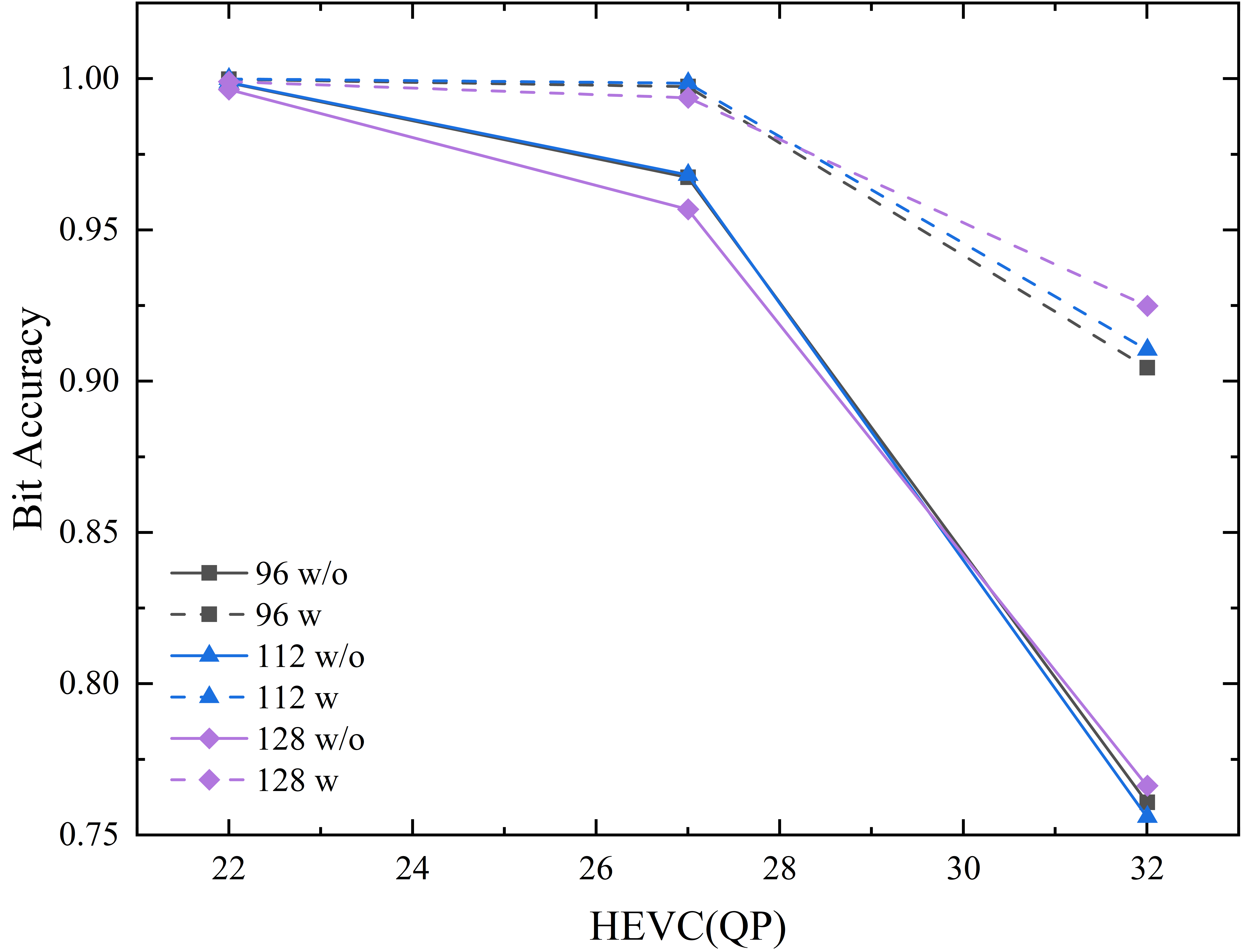}%
\label{WO}}
\hfil
\subfloat[\textrm{Regular rectangular watermark}]{\includegraphics[width=3in]{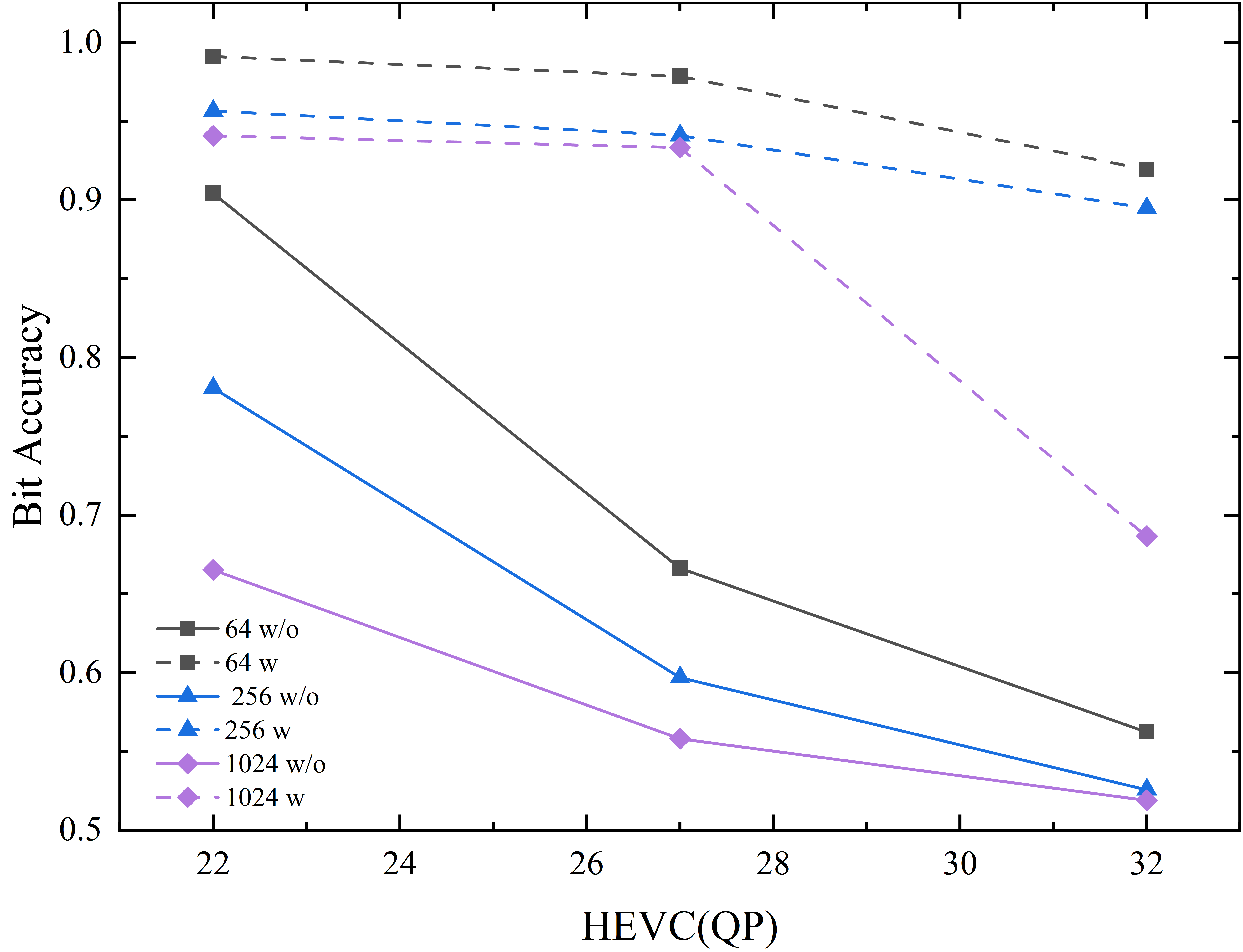}%
\label{WO_2}}
\caption{{Accuracy of DINVMark with and without Distortion layer under HEVC compression attacks}}
\label{merge2}
\end{figure*}

$\textbf{1) Effect of Noise layer:}$
To validate the effectiveness of the distortion layer, we conduct with and without the distortion layer for the network, and HEVC compression is applied at different QP to calculate respective ACC. From Fig. \ref{merge2}, it can be observed that whether embedding regular or irregular watermark, adding a distortion layer to the network that simulates HEVC compression helps to enhance the robustness of the watermark, especially when the QP is 32. This effect is more obvious under these conditions. By comparing Fig. \ref{merge2}, it can be found that in the case of embedding 64-bit watermark without the distortion layer and with QP=32, ACC is lower than that of embedding larger capacity irregular watermark. To validate the effectiveness of the distortion layer, the experiment maintains a similar PSNR for embedding 64-bit watermark with and without the distortion layer, both above 40dB, while the PSNR for embedding higher capacity irregular watermark is 37dB. Therefore, the lower PSNR conditions of the irregular watermark result in better robustness under high compression intensity.

To further analyze why the distortion layer is effective, it is observed that HEVC compression primarily targets the high-frequency components of a video. The distortion layer simulates this process by adding minimally altered information to the low-frequency regions of the video, while using more complex function mappings in the high-frequency regions to learn the specific compression behavior of HEVC on high-frequency content. This design effectively captures the variation characteristics introduced during compression, thereby enhancing the robustness of the watermark against HEVC compression.

\begin{table}[htbp]
\centering
\caption{{The video quality of DINVMark with different Number of block}}
\label{Number of block}
\begin{tabular}{cccc}
\toprule
Number of blocks & PSNR(dB)$\uparrow$ & LPIPS$\downarrow$ & ACC(\%)$\uparrow$\\
\midrule
12 & {36.62} & {0.0606} & 99.98\\
16 & {\textbf{37.50}} & {\underline{0.0241}} & 99.98\\
20 & {\underline{36.74}} & {\textbf{0.0234}} & 99.98\\
\bottomrule
\end{tabular}
\end{table}

$\textbf{2) Influence of the number of invertible blocks:}$
As previously mentioned, the INN is composed of multiple INBs. To assess the impact of varying numbers of INBs on network performance, this paper construct networks with 12, 16, and 20 INBs and evaluate their performance to analyze the influence of the number of INBs on the overall network performance. As shown in Table \ref{Number of block}, with ACC stabilizing at 99.98\%, the model achieved the best performance in terms of PSNR when using 16 INBs. Meanwhile, the LPIPS showed relatively close performance across different INB configurations, indicating that in these configurations, the model has little impact on the perceived video quality. Considering all evaluation metrics, the number of INBs in the network structure is set to 16.

\begin{table}[htbp]
\centering
\caption{{Comparison in terms of model parameters and floating point operations (FLOPs)}}
\label{flops}
\begin{tabular}{lcc}
\toprule
Framework & Param. (M) & FLOPs (G) \\
\midrule
HiDDeN[14] & \textbf{0.53} & \textbf{35} \\
{DVMark[19]} & {21.04} & {2361} \\
REVMark[20] & 7.42 & 157 \\
DINVMark(ours) & \underline{1.38} & \underline{54} \\
\bottomrule
\end{tabular}
\end{table}

\subsection{Comparison in Computational Cost}

{To assess the computational cost, HiDDeN, DVMark, REVMark and DINVMark are compared in terms of model parameters and floating point operations (FLOPs). It is worth noting that DT-CWT was not included in the comparison because it does not employ neural networks.} Since in practical application scenarios, only the encoder and decoder are needed, model parameters and FLOPs for the encoder and decoder were calculated. As shown in Table \ref{flops}, it can be observed that DINVMark requires less model parameters and computation in terms of practical applications and is similar to HiDDeN, which has the lowest computational effort.

\section{Conclusion}

This paper designed an end-to-end lightweight deep invertible network. Initially, an INN is introduced as the main structure for video watermark embedding and extraction. This structure ensures effective watermark embedding while reducing network parameters and computational load. Additionally, a differentiable distortion layer that simulates HEVC compression is designed for end-to-end training, which can effectively guide the watermark into more suitable video regions to resist HEVC video compression. Experiments have shown that compared to existing neural network-based video watermarking schemes, DINVMark can achieve fewer model parameters and less computational effort, while ensuring that the watermarked video maintains higher quality and effectively extracting the watermark from the video. {Certainly, The limitation of this scheme is that the video watermark is not sufficiently robust against cropping attacks. This vulnerability arises from the inherent characteristics of the model and the fact that watermark embedding occurs in the spatial domain. Consequently, intensive cropping attacks can remove specific segments containing watermark information, resulting in its degradation.} Additionally, our scheme has not yet been specifically evaluated under transcoding attacks, as our primary focus has been on enhancing robustness against common compression and noise-related distortions.Thus, this area thus remains a target for improvement in future work.

\bibliographystyle{IEEEtran}
\bibliography{ref}

\begin{thebibliography}{10}
\providecommand{\url}[1]{#1}
\csname url@samestyle\endcsname
\providecommand{\newblock}{\relax}
\providecommand{\bibinfo}[2]{#2}
\providecommand{\BIBentrySTDinterwordspacing}{\spaceskip=0pt\relax}
\providecommand{\BIBentryALTinterwordstretchfactor}{4}
\providecommand{\BIBentryALTinterwordspacing}{\spaceskip=\fontdimen2\font plus
\BIBentryALTinterwordstretchfactor\fontdimen3\font minus
  \fontdimen4\font\relax}
\providecommand{\BIBforeignlanguage}[2]{{%
\expandafter\ifx\csname l@#1\endcsname\relax
\typeout{** WARNING: IEEEtran.bst: No hyphenation pattern has been}%
\typeout{** loaded for the language `#1'. Using the pattern for}%
\typeout{** the default language instead.}%
\else
\language=\csname l@#1\endcsname
\fi
#2}}
\providecommand{\BIBdecl}{\relax}
\BIBdecl

\bibitem{ch1}
R.~Ch., V.~Yadlapalli, S.~S. Sk, G.~T. Reddy, and S.~Kautish, ``Robust
  steganographic framework for securing sensitive healthcare data of
  telemedicine using convolutional neural network,'' \emph{CAAI Transactions on
  Intelligence Technology}, 2024, doi:{\color{blue}
  \href{https://doi.org/10.1049/cit2.12319} {10.1049/cit2.12319}}.

\bibitem{ch2}
C.~Rupa, R.~P. Malleswari, S.~A. Sultana, M.~Abbas, and A.~K. Sahu, ``Data
  privacy protection using lucas series based hybrid reversible watermarking
  approach,'' \emph{IEEE Access}, vol.~12, pp. 134\,578--134\,593, 2024.

\bibitem{01}
S.~He, D.~Xu, L.~Yang, and W.~Liang, ``Adaptive hevc video steganography with
  high performance based on attention-net and pu partition modes,'' \emph{IEEE
  Transactions on Multimedia}, vol.~26, pp. 687--700, 2024.

\bibitem{02}
L.~Yang, D.~Xu, J.~Qian, and R.~Wang, ``Quad-tree structure-preserving adaptive
  steganography for hevc,'' \emph{IEEE Transactions on Multimedia}, vol.~26,
  pp. 8625--8638, 2024.

\bibitem{03}
L.~Yang, R.~Wang, D.~Xu, L.~Dong, and S.~He, ``Centralized error
  distribution-preserving adaptive steganography for hevc,'' \emph{IEEE
  Transactions on Multimedia}, vol.~26, pp. 4255--4270, 2024.

\bibitem{04}
T.~Tokar, T.~Kanocz, and D.~Levicky, ``Digital watermarking of uncompressed
  video in spatial domain,'' in \emph{2009 19th International Conference
  Radioelektronika}, Bratislava, Slovakia, 2009, pp. 319--322.

\bibitem{05}
M.~Asikuzzaman, M.~J. Alam, A.~J. Lambert, and M.~R. Pickering, ``Imperceptible
  and robust blind video watermarking using chrominance embedding: A set of
  approaches in the dt cwt domain,'' \emph{IEEE Transactions on Information
  Forensics and Security}, vol.~9, no.~9, pp. 1502--1517, 2014.

\bibitem{06}
W.~Huan, S.~Li, Z.~Qian, and X.~Zhang, ``Exploring stable coefficients on joint
  sub-bands for robust video watermarking in dt cwt domain,'' \emph{IEEE
  Transactions on Circuits and Systems for Video Technology}, vol.~32, no.~4,
  pp. 1955--1965, 2022.

\bibitem{07}
M.~Asikuzzaman, M.~J. Alam, A.~J. Lambert, and M.~R. Pickering, ``Robust dt
  cwt-based dibr 3d video watermarking using chrominance embedding,''
  \emph{IEEE Transactions on Multimedia}, vol.~18, no.~9, pp. 1733--1748, 2016.

\bibitem{08}
H.~Mareen, J.~De~Praeter, G.~Van~Wallendael, and P.~Lambert, ``A scalable
  architecture for uncompressed-domain watermarked videos,'' \emph{IEEE
  Transactions on Information Forensics and Security}, vol.~14, no.~6, pp.
  1432--1444, 2019.

\bibitem{09}
T.~Stütz, F.~Autrusseau, and A.~Uhl, ``Non-blind structure-preserving
  substitution watermarking of h.264/cavlc inter-frames,'' \emph{IEEE
  Transactions on Multimedia}, vol.~16, no.~5, pp. 1337--1349, 2014.

\bibitem{10}
A.~Mansouri, A.~M. Aznaveh, F.~Torkamani-Azar, and F.~Kurugollu, ``A low
  complexity video watermarking in h.264 compressed domain,'' \emph{IEEE
  Transactions on Information Forensics and Security}, vol.~5, no.~4, pp.
  649--657, 2010.

\bibitem{11}
H.~Mareen, J.~De~Praeter, G.~Van~Wallendael, and P.~Lambert, ``A novel video
  watermarking approach based on implicit distortions,'' \emph{IEEE
  Transactions on Consumer Electronics}, vol.~64, no.~3, pp. 250--258, 2018.

\bibitem{hidden}
J.~Zhu, R.~Kaplan, J.~Johnson, and L.~Fei-Fei, ``Hidden: Hiding data with deep
  networks,'' in \emph{Computer Vision -- ECCV 2018}, Cham, 2018, pp. 682--697.

\bibitem{StegaStamp}
M.~Tancik, B.~Mildenhall, and R.~Ng, ``Stegastamp: Invisible hyperlinks in
  physical photographs,'' in \emph{2020 IEEE/CVF Conference on Computer Vision
  and Pattern Recognition (CVPR)}, Seattle, WA, USA, 2020, pp. 2114--2123.

\bibitem{redmark}
M.~Ahmadi, A.~Norouzi, N.~Karimi, S.~Samavi, and A.~Emami, ``Redmark: Framework
  for residual diffusion watermarking based on deep networks,'' \emph{Expert
  Systems with Applications}, vol. 146, p. 113157, 2020.

\bibitem{211}
S.~Chen, A.~Malik, X.~Zhang, G.~Feng, and H.~Wu, ``A fast method for robust
  video watermarking based on zernike moments,'' \emph{IEEE Transactions on
  Circuits and Systems for Video Technology}, vol.~33, no.~12, pp. 7342--7353,
  2023.

\bibitem{221}
K.~A. Zhang, L.~Xu, A.~Cuesta-Infante, and K.~Veeramachaneni, ``Robust
  invisible video watermarking with attention,'' \emph{arXiv preprint
  arXiv:1909.01285}, 2019.

\bibitem{222}
X.~Luo, Y.~Li, H.~Chang, C.~Liu, P.~Milanfar, and F.~Yang, ``Dvmark: a deep
  multiscale framework for video watermarking,'' \emph{IEEE Transactions on
  Image Processing}, 2023, doi:{\color{blue}
  \href{https://doi.org/10.1109/TIP.2023.3251737} {10.1109/TIP.2023.3251737}}.

\bibitem{223}
Y.~Zhang, J.~Ni, W.~Su, and X.~Liao, ``A novel deep video watermarking
  framework with enhanced robustness to h. 264/avc compression,'' in
  \emph{Proceedings of the 31st ACM International Conference on Multimedia},
  New York, NY, USA, 2023, pp. 8095--8104.

\bibitem{231}
L.~Dinh, D.~Krueger, and Y.~Bengio, ``Nice: Non-linear independent components
  estimation,'' \emph{arXiv preprint arXiv:1410.8516}, 2014.

\bibitem{232}
L.~Dinh, J.~Sohl-Dickstein, and S.~Bengio, ``Density estimation using real
  nvp,'' \emph{arXiv preprint arXiv:1605.08803}, 2016.

\bibitem{2331}
M.~Xiao, S.~Zheng, C.~Liu, Y.~Wang, D.~He, G.~Ke, J.~Bian, Z.~Lin, and T.-Y.
  Liu, ``Invertible image rescaling,'' in \emph{European Conference on Computer
  Vision}, Glasgow’s Scottish Event Campus, 2020.

\bibitem{2341}
J.~Liang, K.~Zhang, S.~Gu, L.~V. Gool, and R.~Timofte, ``Flow-based kernel
  prior with application to blind super-resolution,'' in \emph{2021 IEEE/CVF
  Conference on Computer Vision and Pattern Recognition (CVPR)}, Nashville, TN,
  USA, 2021, pp. 10\,596--10\,605.

\bibitem{2342}
Y.~Liu, Z.~Qin, S.~Anwar, P.~Ji, D.~Kim, S.~Caldwell, and T.~Gedeon,
  ``Invertible denoising network: A light solution for real noise removal,'' in
  \emph{2021 IEEE/CVF Conference on Computer Vision and Pattern Recognition
  (CVPR)}, Nashville, TN, USA, 2021, pp. 13\,360--13\,369.

\bibitem{2351}
T.~F. van~der Ouderaa and D.~E. Worrall, ``Reversible gans for memory-efficient
  image-to-image translation,'' in \emph{2019 IEEE/CVF Conference on Computer
  Vision and Pattern Recognition(CVPR)}, Long Beach,CA, USA, 2019, pp.
  4720--4728.

\bibitem{2361}
J.~Jing, X.~Deng, M.~Xu, J.~Wang, and Z.~Guan, ``Hinet: Deep image hiding by
  invertible network,'' in \emph{2021 IEEE/CVF International Conference on
  Computer Vision (ICCV)}, Montreal, QC, Canada, 2021, pp. 4713--4722.

\bibitem{2362}
S.-P. Lu, R.~Wang, T.~Zhong, and P.~L. Rosin, ``Large-capacity image
  steganography based on invertible neural networks,'' in \emph{2021 IEEE/CVF
  Conference on Computer Vision and Pattern Recognition (CVPR)}, Nashville, TN,
  USA, 2021, pp. 10\,811--10\,820.

\bibitem{2363}
Y.~Xu, C.~Mou, Y.~Hu, J.~Xie, and J.~Zhang, ``Robust invertible image
  steganography,'' in \emph{2022 IEEE/CVF Conference on Computer Vision and
  Pattern Recognition (CVPR)}, New Orleans, LA, USA, 2022, pp. 7865--7874.

\bibitem{2364}
X.~Shen, H.~Yao, S.~Tan, and C.~Qin, ``Vhnet: A video hiding network with
  robustness to video coding,'' \emph{Journal of Information Security and
  Applications}, vol.~75, p. 103515, 2023.

\bibitem{2365}
C.~Mou, Y.~Xu, J.~Song, C.~Zhao, B.~Ghanem, and J.~Zhang, ``Large-capacity and
  flexible video steganography via invertible neural network,'' in \emph{2023
  IEEE/CVF Conference on Computer Vision and Pattern Recognition (CVPR)},
  Vancouver, BC, Canada, 2023, pp. 22\,606--22\,615.

\bibitem{2371}
H.-B. Xu, R.~Wang, J.~Wei, and S.-P. Lu, ``A compact neural network-based
  algorithm for robust image watermarking,'' \emph{arXiv preprint
  arXiv:2112.13491}, 2021.

\bibitem{2372}
R.~Ma, M.~Guo, Y.~Hou, F.~Yang, Y.~Li, H.~Jia, and X.~Xie, ``Towards blind
  watermarking: Combining invertible and non-invertible mechanisms,'' in
  \emph{Proceedings of the 30th ACM International Conference on Multimedia},
  Lisbon, Portugal, 2022, pp. 1532--1542.

\bibitem{2373}
H.~Fang, Y.~Qiu, K.~Chen, J.~Zhang, W.~Zhang, and E.-C. Chang, ``Flow-based
  robust watermarking with invertible noise layer for black-box distortions,''
  in \emph{Proceedings of the AAAI conference on artificial intelligence},
  vol.~37, no.~4, Washington, DC, USA, 2023, pp. 5054--5061.

\bibitem{tgan}
M.~Saito, S.~Saito, M.~Koyama, and S.~Kobayashi, ``Train sparsely, generate
  densely: Memory-efficient unsupervised training of high-resolution temporal
  gan,'' \emph{International Journal of Computer Vision}, vol. 128, no.~10, pp.
  2586--2606, 2020.

\bibitem{tlp}
M.~Chu, Y.~Xie, J.~Mayer, L.~Leal-Taix\'{e}, and N.~Thuerey, ``Learning
  temporal coherence via self-supervision for gan-based video generation,''
  \emph{ACM Transactions on Graphics (TOG)}, vol.~39, no.~4, Aug. 2020.

\end{thebibliography}

\begin{IEEEbiography}[{\includegraphics[width=1in,height=1.25in,clip,keepaspectratio]{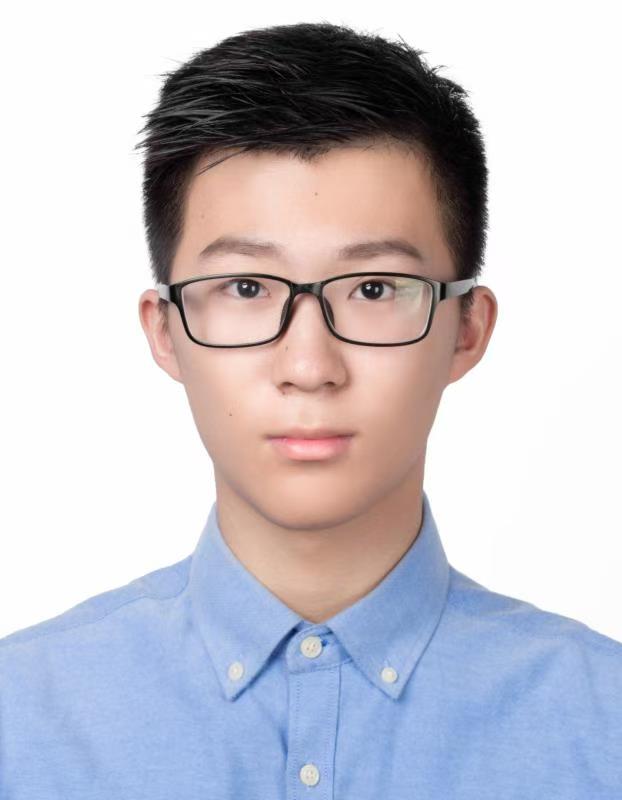}}]{Jianbin Ji}He received the B.E. degree in the School of Cyber Science and Engineering, Ningbo University of Technology, Ningbo, China, in 2023. He is currently pursuing the M.E. degree with the Faculty of Electrical Engineering and Computer Science, Ningbo University, Ningbo, China.

His research interests include video watermarking.

\end{IEEEbiography}

\begin{IEEEbiography}[{\includegraphics[width=1in,height=1.25in,clip,keepaspectratio]{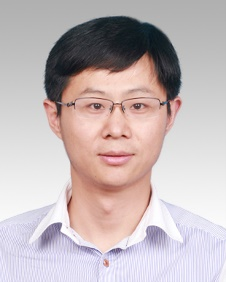}}]{Dawen Xu}(Senior Member, IEEE) He received the M.S. degree in communication and information system from Ningbo University, Ningbo, China, in 2005, and the Ph.D. degree in computer applied technology from Tongji University, Shanghai, China, in 2011. He was a Visiting Scholar with New Jersey Institute of Technology, Newark, NJ, USA, from 2012 to 2013. He is currently a Professor with School of Cyber Science and Engineering, Ningbo University of Technology, Ningbo, China.

His research interests include digital watermarking and information hiding, steganography and steganalysis, and signal processing in the encrypted domain. He has served as a Technical Paper Reviewer for IEEE conferences, journal, and magazines.

\end{IEEEbiography}

\begin{IEEEbiography}[{\includegraphics[width=1in,height=1.25in,clip,keepaspectratio]{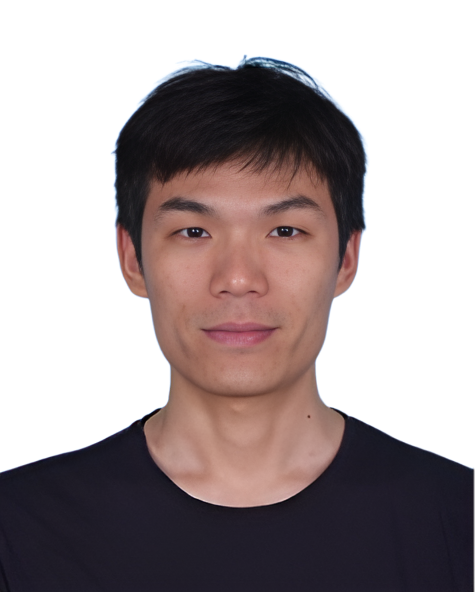}}]{Dong Li}(Member, IEEE) He received the B.E. degree from Chongqing University, Chongqing, China, in 2012, and the M.S. and Ph.D. degrees from the University of Macau, Macau, China, in 2014 and 2018, respectively. He is currently an Associate Professor with the Faculty of electrical engineering and computer science, Department of Computer Science, Ningbo University, Ningbo, China.

His research interests include statistical image modeling and processing, multimedia security and forensic, and machine learning.

\end{IEEEbiography}

\begin{IEEEbiography}[{\includegraphics[width=1in,height=1.25in,clip,keepaspectratio]{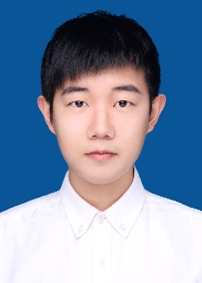}}]{Lin Yang}He was born in Ningbo, China, in 1997. He received M.S. degree from the Faculty of Electrical Engineering and Computer Science, Ningbo University, Ningbo, China. He is currently pursuing the Ph.D. degree with Professor Jiangbo Qian and Professor Dawen Xu in the Faculty of Electrical Engineering and Computer Science, Ningbo University, Ningbo, China.

His research interests include video steganography and steganalysis.

\end{IEEEbiography}

\vspace{-40\baselineskip}
\begin{IEEEbiography}[{\includegraphics[width=1in,height=1.25in,clip,keepaspectratio]{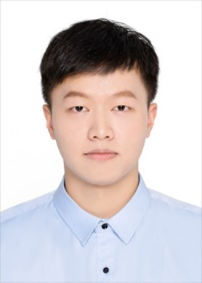}}]{Songhan He}He was born in Ningbo, China, in 1998. He received the M.S. degree from the Faculty of Electrical Engineering and Computer Science, Ningbo University, Ningbo, China. He is currently working toward the Ph.D. degree with Professor Dawen Xu with the Faculty of Electrical Engineering and Computer Science, Ningbo University, Ningbo.

His research interests include video steganography, steganalysis and watermarking.

\end{IEEEbiography}

\end{document}